\documentclass{emulateapj}
\usepackage{apjfonts}

\begin{document}

\lefthead{Chandra Observations of IGR Sources}
\righthead{Tomsick et al.}

\submitted{Accepted by the Astrophysical Journal}

\def\lsim{\mathrel{\lower .85ex\hbox{\rlap{$\sim$}\raise
.95ex\hbox{$<$} }}}
\def\gsim{\mathrel{\lower .80ex\hbox{\rlap{$\sim$}\raise
.90ex\hbox{$>$} }}}

\title{{\em Chandra} Localizations and Spectra of {\em INTEGRAL}
Sources in the Galactic Plane:\\  The Cycle 9 Sample}

\author{John A. Tomsick\altaffilmark{1},
Sylvain Chaty\altaffilmark{2},
Jerome Rodriguez\altaffilmark{2},
Roland Walter\altaffilmark{3},
Philip Kaaret\altaffilmark{4}}

\altaffiltext{1}{Space Sciences Laboratory, 7 Gauss Way, 
University of California, Berkeley, CA 94720-7450, USA
(e-mail: jtomsick@ssl.berkeley.edu)}

\altaffiltext{2}{AIM - Astrophysique Interactions Multi-\'echelles
(UMR 7158 CEA/CNRS/Universit\'e Paris 7 Denis Diderot),
CEA Saclay, DSM/IRFU/Service d'Astrophysique, B\^at. 709,
L'Orme des Merisiers, FR-91 191 Gif-sur-Yvette Cedex, France}

\altaffiltext{3}{INTEGRAL Science Data Centre, Observatoire
de Gen\`eve, Universit\'e de Gen\`eve, Chemin d'Ecogia, 16, 
1290 Versoix, Switzerland}

\altaffiltext{4}{Department of Physics and Astronomy, University of
Iowa, Iowa City, IA 52242, USA}

\begin{abstract}

We report on 0.3--10 keV X-ray observations by the {\em Chandra X-ray 
Observatory} of the fields of 22 sources that were discovered as hard
X-ray (20--100 keV) sources by the {\em INTEGRAL} satellite (``IGR'' 
sources).  The observations were made during {\em Chandra}'s 9th observing 
cycle, and their purpose is to localize the sources and to measure their 
soft X-ray spectra in order to determine the nature of the sources.  We 
find very likely {\em Chandra} counterparts for 18 of the 22 sources.  
We discuss the implications for each source, considering previous results 
and new optical or IR identifications, and we identify or suggest 
identifications for the nature of 16 of the sources.  Two of the sources, 
IGR~J14003--6326 and IGR~J17448--3232, are extended on arcminute scales.  
We identify the former as a pulsar wind nebula (PWN) with a surrounding 
supernova remnant (SNR) and the latter as a SNR.  In the group of 242 IGR 
sources, there is only one other source that has previously been identified 
as a SNR.  Seven of the sources are definite or candidate High-Mass X-ray 
Binaries (HMXBs).  We confirm a previous identification of IGR~J14331--6112 
as an HMXB, and based on combinations of hard X-ray spectra, inferred 
distances and X-ray luminosities, and/or column density variations, we 
suggest that IGR~J17404--3655, IGR~J16287--5021, IGR~J17354--3255, 
IGR~J17507--2647, IGR~J17586--2129, and IGR~J13186--6257 are candidate 
HMXBs.  Our results indicate or confirm that IGR~J19267+1325, 
IGR~J18173--2509, and IGR~J18308--1232 are Cataclysmic Variables (CVs), 
and we suggest that IGR~J15529--5029 may also be a CV.  We confirm that 
IGR~J14471--6414 is an Active Galactic Nucleus (AGN), and we also suggest 
that IGR~J19443+2117 and IGR~J18485--0047 may be AGN.  Finally, we found
{\em Chandra} counterparts for IGR~J11098--6457 and IGR~J18134--1636, but
more information is required to determine the nature of these two sources.

\end{abstract}


\section{Introduction}

The {\em International Gamma-Ray Astrophysics Satellite (INTEGRAL)} 
\citep{winkler03} has been in operation for more than six years and has 
discovered a large number of new hard X-ray sources at energies $>$20~keV.  
This bandpass is above the range where most types of sources can emit thermal 
blackbody radiation.  Thus, {\em INTEGRAL} provides a window on the non-thermal 
emission that comes from Galactic sources (e.g., accreting black holes, neutron 
stars, and magnetized white dwarfs) as well as extragalactic Active Galactic 
Nuclei (AGN).  With large field-of-view (FOV) coded aperture mask instruments 
\citep[e.g., a $29^{\circ}\times 29^{\circ}$ partially coded FOV for IBIS,][]{lebrun03,ubertini03}, 
{\em INTEGRAL} has observed most of the sky thus far during its mission, but 
its sky coverage has not been uniform, with more time being concentrated near 
the Galactic plane.  While IBIS has provided a step forward in hard X-ray 
angular resolution, it localizes 20--40~keV sources to 1--5$^{\prime}$, which is 
normally not sufficient to identify an optical or infrared counterpart, leaving
the nature of many of the sources uncertain.  Thus, throughout the {\em INTEGRAL} 
mission, follow-up observations have been made with soft X-ray ($\sim$0.5--10 keV)
telescopes ({\em Chandra}, {\em XMM-Newton}, and {\em Swift}) to localize the 
{\em INTEGRAL} sources and determine their nature.

As of 2009 March, {\em INTEGRAL} had detected nearly 600 sources, and 242 of 
these are IGR sources\footnote{See http://isdc.unige.ch/$\sim$rodrigue/html/igrsources.html
for a continually updated list of IGR sources.}, being either newly discovered 
by {\em INTEGRAL} or not previously detected in the $\sim$20--100~keV band
\cite[see][for more on the criterion for the IGR classification]{bodaghee07}.  
The source types of the 242 IGR sources are dominated by 79 AGN (73 with firm 
identifications), 67 X-ray binaries (47 with firm identifications), and 15 Cataclysmic 
Variables (CVs).  There are small numbers of IGR sources in a few other categories 
(Symbiotic stars, active stars, an Anomalous X-ray Pulsar, and a Supernova Remnant), 
and the remaining 72 IGR sources are unclassified.  Along with the soft X-ray 
localizations \citep[e.g.,][]{tomsick06,tomsick08}, the steady increase in the numbers 
of classified IGR sources depends on ground-based optical and IR follow-up observations 
\citep[e.g.,][]{masetti6,masetti7,chaty08,butler09}, which are key to obtaining firm 
identifications.

One of the most interesting categories of IGR sources are the High-Mass X-ray
Binaries (HMXBs), for which {\em INTEGRAL} has uncovered two new (and not 
entirely independent) classes: The obscured HMXBs \citep{walter03,mg03,walter06} 
and the Supergiant Fast X-ray Transients \citep[SFXTs,][]{negueruela06,sguera06,wz07}.
The obscured HMXBs have large and, in some cases, variable levels of intrinsic 
(i.e., local) absorption with $N_{\rm H}\sim 10^{23-24}$ cm$^{-2}$.  There is evidence
in several cases that a strong stellar equatorial outflow is responsible for the 
absorption, including the detection of P~Cygni profiles \citep{fc04,chaty08}, 
providing direct evidence for a wind, as well as excess emission in the mid-infrared
\citep{rahoui08}.  The strong winds and high absorption explain why {\em INTEGRAL}'s
hard X-ray imaging was necessary to detect these sources, and it has been 
suggested that these obscured HMXBs are in a previously unstudied phase of 
HMXB evolution \citep{lommen05}.  With the discovery of SFXTs, {\em INTEGRAL} found
a new type of HMXB behavior where sources undergo high-amplitude (orders of magnitude)
hard X-ray flares that only last for time scales of hours \citep{intzand05,sguera06}.
In part, this is thought to be caused by clumps in the supergiant stellar winds 
\citep{wz07}, but it may also be related to the interaction between the wind and 
a strong neutron star magnetic field \citep{bfs08}.

To identify the nature of more of these sources, we have been using {\em Chandra}
to study the unclassified IGR sources near the Galactic plane.  The primary purpose
of the observations is to localize the sources to allow for optical or IR follow-up.
While X-ray positions with an accuracy of several arcseconds (as can be given by
{\em Swift} or {\em XMM-Newton}) can lead to correct optical or IR identifications, 
in many cases (and especially in the crowded regions of the Galactic plane), the
sub-arcsecond {\em Chandra} positions are crucial for obtaining definitive 
identifications.  A second important use for the {\em Chandra} observations is 
to measure the soft X-ray energy spectrum, providing information about source 
hardness and levels of absorption that help to determine the nature of the source.
Our {\em Chandra} program began in 2005 with the identification of 4 IGR 
sources \citep{tomsick06}, and we subsequently used {\em Chandra} observations 
taken between 2007 and 2008 to obtain identifications for 12 more IGR sources 
\citep{tomsick08a}.  In this work, we report on 22 {\em Chandra} observations 
taken during {\em Chandra}'s 9th observing cycle between late-2007 and early-2009.

\section{{\em Chandra} Observations}

For {\em Chandra} follow-up observations, we chose sources from the 27th
{\em INTEGRAL} General Reference Catalog\footnote{See http://isdc.unige.ch/Data/cat.}
that were detected in the 20--40 keV band as of 2007 March.  The list that we 
produced includes IGR sources within $5^{\circ}$ of the Galactic plane, and we used 
information from \cite{bird06} and \cite{bodaghee07} to select sources that were not 
known to be transient and whose nature was unknown at the time.  Ultimately, we chose 
the 22 sources listed in Table~\ref{tab:obs} for $\sim$5~ks ``snapshot'' observations 
with {\em Chandra} to provide precise X-ray localizations as well as information about 
their soft X-ray spectra.

\begin{table*}
\caption{{\em Chandra} Observations\label{tab:obs}}
\begin{minipage}{\linewidth}
\begin{center}
\begin{tabular}{cccclc} \hline \hline
IGR Name & ObsID & $l$\footnote{Galactic longitude in degrees.} & $b$\footnote{Galactic latitude in degrees.} & Start Time & Exposure Time (s)\\ \hline \hline
J07295--1329 & 9061 & 228.97 &  +2.26 & 2008 Feb  11,  0.55 h UT & 5093\\
J09485--4726 & 9068 & 273.84 &  +4.84 & 2007 Nov  30, 19.98 h UT & 5055\\
J11098--6457 & 9066 & 292.43 & --4.17 & 2008 Sept 12, 14.66 h UT & 5109\\
J13186--6257 & 9049 & 306.02 & --0.24 & 2008 Sept 11, 22.14 h UT & 5061\\
J14003--6326 & 9058 & 310.57 & --1.61 & 2008 June 29, 12.13 h UT & 5077\\
J14331--6112 & 9053 & 314.90 & --0.72 & 2008 Jan   5,  4.39 h UT & 4888\\
J14471--6414 & 9065 & 315.00 & --4.15 & 2007 Dec  27,  7.83 h UT & 5112\\
J15529--5029 & 9062 & 329.89 &  +2.63 & 2008 Jan   7,  8.58 h UT & 5071\\
J16287--5021 & 9054 & 334.16 & --1.13 & 2008 Jan  28, 10.84 h UT & 4935\\
J17354--3255 & 9050 & 355.44 & --0.26 & 2009 Feb   6,  7.15 h UT & 4692\\
J17404--3655 & 9063 & 352.64 & --3.27 & 2008 July 16, 21.02 h UT & 4894\\
J17448--3232 & 9059 & 356.84 & --1.76 & 2008 Nov   2,  2.15 h UT & 4692\\
J17461--2204 & 9064 &   5.94 &  +3.48 & 2009 Feb   6,  5.27 h UT & 4888\\
J17487--3124 & 9060 & 358.25 & --1.83 & 2008 Nov   2,  0.33 h UT & 4888\\
J17507--2647 & 9048 &   2.42 &  +0.15 & 2009 Feb   6,  8.79 h UT & 4698\\
J17586--2129 & 9056 &   8.05 &  +1.35 & 2008 Oct  30,  3.38 h UT & 4888\\
J18134--1636 & 9052 &  13.88 &  +0.61 & 2008 Feb  16,  9.52 h UT & 4891\\
J18173--2509 & 9067 &   6.79 & --4.26 & 2008 May   6,  1.15 h UT & 4894\\
J18308--1232 & 9055 &  19.45 & --1.19 & 2008 Apr  27, 16.69 h UT & 4698\\
J18485--0047 & 9051 &  31.90 &  +0.31 & 2008 Apr  27, 14.97 h UT & 4695\\
J19267+1325  & 9075 &  48.88 & --1.53 & 2008 Feb  27,  2.46 h UT & 4698\\
J19443+2117  & 9057 &  57.79 & --1.37 & 2008 Feb  27,  0.68 h UT & 4888\\ \hline
\end{tabular}
\end{center}
\end{minipage}
\end{table*}

As listed in Table~\ref{tab:obs}, the observations were made between 2007 November
and 2009 February, and we used the Advanced CCD Imaging Spectrometer 
\citep[ACIS,][]{garmire03}.  The 90\% confidence {\em INTEGRAL} error circles range 
from $2^{\prime}.2$ to $5^{\prime}.4$, leading us to use the $16.9\times 16.9$ arcmin$^{2}$
field-of-view (FOV) of the ACIS-I instrument.  Our analysis of the ACIS data began
by downloading the most recent versions of the data products as of 2009 
January-February.  We started with the ``level 1'' data files, and these were
processed at the {\em Chandra} X-Ray Center with pipeline (``ASCDS'') versions
7.6.11.2 to 7.6.11.9.  We performed further processing with the {\em Chandra}
Interactive Analysis of Observations (CIAO) version 4.1.1 (except for one part of
the analysis as mentioned below) software.  Also, we used version 4.1.1 of the
Calibration Data Base (CALDB).  We used the CIAO routine {\ttfamily acis\_process\_events}
to produce the ``level 2'' event lists that we used to make images and energy
spectra.  While a level 2 file is provided as part of the standard data pipeline, 
we re-produced the file in order to use the most recent time-dependent gain 
information.  With a current version of the CALDB, using the CIAO 4.1.1 default 
parameters when running {\ttfamily acis\_process\_events} incorporates the
most recent calibration files.

\section{Analysis and Results}

\subsection{Search for Sources and X-ray Identifications}

After producing 0.3--10~keV images, we used the CIAO routine {\ttfamily wavdetect} to 
search for X-ray sources on the ACIS-I chips.  Although we originally used the version
of {\ttfamily wavdetect} found in CIAO 4.1.1, we noticed that many spurious detections
(sources with one or two photons) occurred for some of the observations.  However, 
these spurious detections did not occur with CIAO 3.4, and we used that version of
{\ttfamily wavdetect} for the source searches in this paper.  For each observation, 
we searched for sources in unbinned images with $2048\times 2048$ pixels as well as
images binned by a factor of 2 ($1024\times 1024$ pixels) and a factor of 4
($512\times 512$ pixels).  In each case, we set the detection threshold to a level
that would be expected to yield one spurious source ($2.4\times 10^{-7}$, $9.5\times 10^{-7}$, 
and $3.8\times 10^{-6}$ for the three images, respectively).  Thus, in the merged
lists of detected sources, it would not be surprising if a few of the sources in
each field are spurious.  

The average number of sources detected per observation over the entire ACIS-I
FOV is 20, with the minimum being 12 sources and the maximum being 32 sources.  
We used 3 main criteria to determine which of these are likely counterparts to 
the IGR sources.  First, the IGR sources have been detected by {\em INTEGRAL} at 
20--40 keV flux levels of 0.5--2 millicrab \citep{bird06}.  At 1 millicrab (20--40 keV), 
a source with an absorbed power-law spectrum with a photon index of $\Gamma = 1.0$ and 
a column density of $N_{\rm H} = 5\times 10^{22}$ cm$^{-2}$ would yield $\sim$300 ACIS-I 
counts in 5~ks, so we are looking for fairly bright {\em Chandra} sources.  Second, 
the sources should have hard spectra, and third, the sources should be within or 
close to the {\em INTEGRAL} error circles. 

We estimated the hardness of each of the detected {\em Chandra} sources by determining
the number of source counts in the 0.3--2 keV band ($C_{1}$) and the 2--10 keV band 
($C_{2}$).  Using the size of the {\em Chandra} point spread function (PSF) as a guide,
we used extraction radii of $5^{\prime\prime}$ for sources within $4^{\prime}$ of the aimpoint, 
$10^{\prime\prime}$ for sources between $4^{\prime}$ and $7^{\prime}$ from the aimpoint, and
$15^{\prime\prime}$ for sources more than $7^{\prime}$ from the aimpoint.  The background
counts were taken from the largest possible rectangular source-free region, and this
value was scaled to the size of the source extraction region and subtracted off.  
We then calculated the hardness according to $(C_{2}-C_{1})/(C_{2}+C_{1})$, which runs
from --1.0 for the softest sources (all the counts in the 0.3--2 keV band) to +1.0
for the hardest sources (all the counts in the 2--10 keV band).  To deal with the
case where there are zero counts in one of the energy bands, we used the ``Gehrels''
prescription for determining the uncertainties \citep{gehrels86}.

\begin{figure}
\includegraphics[clip, scale=0.45]{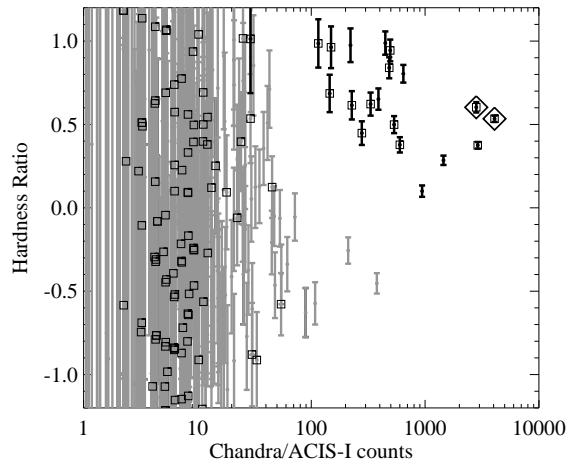}
\caption{Hardness-intensity diagram, including all of the {\em Chandra}/ACIS-I sources detected in
the 22 {\em Chandra} observations.  The intensity is given as the number of ACIS-I counts in the 
0.3--10 keV energy band.  The hardness is given by $(C_{2}-C_{1})/(C_{2}+C_{1})$, where $C_{1}$ is
the number of counts in the 0.3--2 keV band and $C_{2}$ is the number of counts in the 2--10 keV
band.  The squares indicate sources that are inside their respective 90\% confidence {\em INTEGRAL}
error circles.  The sources that we consider as possible {\em Chandra} counterparts to the IGR 
sources are plotted with black points and error bars.  The two sources marked with diamonds are 
extended sources.\label{fig:hi}}
\end{figure}

Figure~\ref{fig:hi} shows a hardness-intensity diagram, including all 434 of the 
sources detected.  For most sources with fewer than $\sim$10 counts, the uncertainties
on the hardnesses are very large.  However, one can see a group of 19 sources with 
more than 100 counts per source and hardnesses greater than zero, and we consider
these as candidate counterparts to the IGR sources.  In addition to the 19 sources, 
we also consider a 20th source that only has 29 counts, but its hardness is 
$1.0\pm 0.3$, and it is well within the {\em INTEGRAL} error circle for 
IGR~J13186--6257.  Thus, we consider these 20 sources as candidate counterparts.
While 18 of these sources are consistent with being point sources, two of the sources 
(marked with diamonds in Figure~\ref{fig:hi}) are clearly extended.  The {\em Chandra} 
positions, numbers of ACIS-I counts in the 0.3--10 keV band, and hardness ratios 
(using the definition given above) for the 20 sources are given in 
Table~\ref{tab:counterparts}.  The 20 sources come from 18 different observations.  
The candidate counterparts for 16 of the IGR sources are unique, and in two cases 
(IGR~J17354--3255 and IGR~J17448--3232), there are 2 possible counterparts.  
In \S$3.3$, we estimate the probability of spurious {\em Chandra}/IGR source 
associations.  For the other 4 IGR sources (IGR~J07295--1329, IGR~J09485--4726, 
IGR~J17461--2204, and IGR~J17487--3124), there are no strong possibilities for 
counterparts.  However, {\em Chandra} sources are detected in each of these 4 fields 
as well, and {\em Chandra} source lists can be found in the on-line tables associated 
with this paper (see Appendix).

\subsection{Spectral Analysis of Point Sources}

\begin{table*}
\caption{{\em Chandra} Candidate Counterparts to IGR Sources\label{tab:counterparts}}
\begin{minipage}{\linewidth}
\begin{center}
\begin{tabular}{cccccc} \hline \hline
CXOU Name & 
$\theta$ (arcminutes)\footnote{The angular distance between the center of the {\em INTEGRAL} error circle, which is also the approximate {\em Chandra} aimpoint, and the source.} & 
{\em Chandra} R.A. (J2000)\footnote{The position uncertainties for these relatively bright sources are dominated by the systematic error, which is $0^{\prime\prime}.64$ at 90\% confidence and $1^{\prime\prime}$ at 99\% confidence \citep{weisskopf05}.} & 
{\em Chandra} Decl. (J2000)$^{\it b}$ & 
ACIS Counts & 
Hardness Ratio\\ \hline
\multicolumn{6}{c}{IGR J11098-6457, {\em INTEGRAL} error radius =  5.5 arcmin\footnote{These are the 90\% confidence errors given in \cite{bird06}.}}\\
J110926.4--650224 &  6.02 & $11^{\rm h}09^{\rm m}26^{\rm s}.43$ & --$65^{\circ}02^{\prime}25^{\prime\prime}\!.0$ & 941.2 &   $0.10\pm 0.03$\\ \hline
\multicolumn{6}{c}{IGR J13186--6257, {\em INTEGRAL} error radius = 3.8 arcmin$^{\it c}$}\\
J131825.0--625815 &  1.94 & $13^{\rm h}18^{\rm m}25^{\rm s}.08$ & --$62^{\circ}58^{\prime}15^{\prime\prime}\!.5$ &  29.3 &   $1.01\pm 0.33$\\ \hline
\multicolumn{6}{c}{IGR J14003--6326, {\em INTEGRAL} error radius = 3.9 arcmin$^{\it c}$}\\
J140045.6--632542\footnote{Extended source.} &  1.47 & $14^{\rm h}00^{\rm m}45^{\rm s}.69$ & --$63^{\circ}25^{\prime}42^{\prime\prime}\!.6$ & 4075.5 &   $0.53\pm 0.02$\\ \hline
\multicolumn{6}{c}{IGR J14331--6112, {\em INTEGRAL} error radius = 3.7 arcmin$^{\it c}$}\\
J143308.3--611539 &  4.02 & $14^{\rm h}33^{\rm m}08^{\rm s}.33$ & --$61^{\circ}15^{\prime}39^{\prime\prime}\!.9$ & 644.1 &   $0.80\pm 0.05$\\ \hline
\multicolumn{6}{c}{IGR J14471--6414, {\em INTEGRAL} error radius = 4.6 arcmin$^{\it c}$}\\
J144628.2--641624 &  1.46 & $14^{\rm h}46^{\rm m}28^{\rm s}.26$ & --$64^{\circ}16^{\prime}24^{\prime\prime}\!.1$ & 600.3 &   $0.38\pm 0.05$\\ \hline
\multicolumn{6}{c}{IGR J15529--5029, {\em INTEGRAL} error radius = 3.9 arcmin$^{\it c}$}\\
J155246.9--502953 &  1.59 & $15^{\rm h}52^{\rm m}46^{\rm s}.92$ & --$50^{\circ}29^{\prime}53^{\prime\prime}\!.4$ & 278.3 &   $0.45\pm 0.07$\\ \hline
\multicolumn{6}{c}{IGR J16287--5021, {\em INTEGRAL} error radius = 4.4 arcmin$^{\it c}$}\\
J162826.8--502239 &  3.12 & $16^{\rm h}28^{\rm m}26^{\rm s}.85$ & --$50^{\circ}22^{\prime}39^{\prime\prime}\!.7$ & 484.3 &   $0.84\pm 0.06$\\ \hline
\multicolumn{6}{c}{IGR J17354--3255, {\em INTEGRAL} error radius = 2.2 arcmin$^{\it c}$}\\
J173518.7--325428 &  1.83 & $17^{\rm h}35^{\rm m}18^{\rm s}.73$ & --$32^{\circ}54^{\prime}28^{\prime\prime}\!.7$ & 145.2 &   $0.69\pm 0.11$\\
J173527.5--325554 &  1.34 & $17^{\rm h}35^{\rm m}27^{\rm s}.59$ & --$32^{\circ}55^{\prime}54^{\prime\prime}\!.4$ & 494.2 &   $0.94\pm 0.07$\\ \hline
\multicolumn{6}{c}{IGR J17404--3655, {\em INTEGRAL} error radius = 3.5 arcmin$^{\it c}$}\\
J174026.8--365537 &  0.84 & $17^{\rm h}40^{\rm m}26^{\rm s}.86$ & --$36^{\circ}55^{\prime}37^{\prime\prime}\!.4$ & 227.3 &   $0.62\pm 0.09$\\ \hline
\multicolumn{6}{c}{IGR J17448--3232, {\em INTEGRAL} error radius = 2.2 arcmin$^{\it c}$}\\
J174453.4--323254$^{\it d}$ &  0.33 & $17^{\rm h}44^{\rm m}53^{\rm s}.44$ & --$32^{\circ}32^{\prime}54^{\prime\prime}\!.1$ & 2816.3 &   $0.60\pm 0.03$\\
J174437.3--323222 &  3.77 & $17^{\rm h}44^{\rm m}37^{\rm s}.34$ & --$32^{\circ}32^{\prime}23^{\prime\prime}\!.0$ &  389.0 &   $0.65\pm 0.06$\\ \hline
\multicolumn{6}{c}{IGR J17507--2647, {\em INTEGRAL} error radius = 2.6 arcmin$^{\it c}$}\\
J175039.4--264436 &  2.97 & $17^{\rm h}50^{\rm m}39^{\rm s}.47$ & --$26^{\circ}44^{\prime}36^{\prime\prime}\!.2$ &  221.1 &   $0.98\pm 0.10$\\ \hline
\multicolumn{6}{c}{IGR J17586--2129, {\em INTEGRAL} error radius = 3.3 arcmin$^{\it c}$}\\
J175834.5--212321 &  3.81 & $17^{\rm h}58^{\rm m}34^{\rm s}.56$ & --$21^{\circ}23^{\prime}21^{\prime\prime}\!.6$ &  447.2 &    $0.99\pm 0.07$\\ \hline
\multicolumn{6}{c}{IGR J18134--1636, {\em INTEGRAL} error radius = 3.8 arcmin$^{\it c}$}\\
J181328.0--163548 &  0.97 & $18^{\rm h}13^{\rm m}28^{\rm s}.03$ & --$16^{\circ}35^{\prime}48^{\prime\prime}\!.5$ & 149.2 &   $0.96\pm 0.13$\\ \hline
\multicolumn{6}{c}{IGR J18173--2509, {\em INTEGRAL} error radius = 2.2 arcmin$^{\it c}$}\\
J181722.1--250842 &  0.86 & $18^{\rm h}17^{\rm m}22^{\rm s}.18$ & --$25^{\circ}08^{\prime}42^{\prime\prime}\!.5$ & 332.2 &   $0.62\pm 0.07$\\ \hline
\multicolumn{6}{c}{IGR J18308--1232, {\em INTEGRAL} error radius = 3.3 arcmin$^{\it c}$}\\
J183049.9--123219 &  0.83 & $18^{\rm h}30^{\rm m}49^{\rm s}.95$ & --$12^{\circ}32^{\prime}19^{\prime\prime}\!.1$ & 534.3 &   $0.50\pm 0.05$\\ \hline
\multicolumn{6}{c}{IGR J18485--0047, {\em INTEGRAL} error radius = 3.4 arcmin$^{\it c}$}\\
J184825.4--004635 &  0.55 & $18^{\rm h}48^{\rm m}25^{\rm s}.47$ & --$00^{\circ}46^{\prime}35^{\prime\prime}\!.2$ & 115.3 &   $0.99\pm 0.15$\\ \hline
\multicolumn{6}{c}{IGR J19267+1325, {\em INTEGRAL} error radius = 3.7 arcmin$^{\it c}$}\\
J192626.9+132205  &  4.79 & $19^{\rm h}26^{\rm m}26^{\rm s}.99$ &  +$13^{\circ}22^{\prime}05^{\prime\prime}\!.1$ & 1450.3 &   $0.29\pm 0.03$\\ \hline
\multicolumn{6}{c}{IGR J19443+2117, {\em INTEGRAL} error radius = 4.9 arcmin$^{\it c}$}\\
J194356.2+211823  &  4.88 & $19^{\rm h}43^{\rm m}56^{\rm s}.23$ &  +$21^{\circ}18^{\prime}23^{\prime\prime}\!.6$ & 2902.2 &   $0.38\pm 0.02$\\ \hline
\end{tabular}
\end{center}
\end{minipage}
\end{table*}

We produced {\em Chandra} energy spectra for the 18 point source candidate 
counterparts using the CIAO software routines.  We used the same source 
extraction regions described above for determining count rates and hardnesses.
For the background, we used rectangular source-free regions as described above, 
but they were modified so that they would include counts from only the ACIS-I 
CCD chip that also includes the source.  There are 4 ACIS-I CCD chips, and in 
some cases, the spacecraft dithering causes source counts to be spread across 
multiple CCD chips.  In this case, we put the background region on the CCD chip 
containing the largest number of source counts.  Once the source and background 
regions were determined, we used the CIAO routine {\ttfamily dmextract} to 
produce energy spectra and {\ttfamily mkacisrmf} and {\ttfamily mkarf} to 
produce the response files for the source region.  In the cases where the source 
counts are spread over multiple CCD chips, we used a weight map to produce
weighted response files.

We fitted the 0.3--10~keV ACIS spectra using the XSPEC v12 software.
Initially, we rebinned the spectra to 13 energy bins prior to fitting
with an absorbed power-law model using $\chi^{2}$-minimization in order 
to be able to see if and how the spectra deviate from this basic model.
To account for absorption, we used the photoelectric absorption cross 
sections from \cite{bm92} and elemental abundances from \cite{wam00}, 
which correspond to the estimated abundances for the interstellar medium.
For 12 of the 18 point sources, no pattern in the residuals to the 
absorbed power-law are apparent, and the reduced-$\chi^{2}$ values in 
these cases range from $\chi^{2}_{\nu}$ = 0.22 to 1.5 for 10 degrees of
freedom (dof).  For 5 of the remaining sources (CXOU J144628.2--641624,
CXOU J173527.5--325554, CXOU J181722.1--250842, CXOU J183049.9--123219, 
and CXOU J194356.2+211823), a high energy excess is present above
$\sim$7~keV, which most likely indicates that the spectra are affected
by pile-up.  We confirmed that pile-up is the most likely explanation
by looking at the numbers of counts above 10~keV for all 18 spectra.
Although ACIS has very little sensitivity above 10~keV, these 5 sources
have 12, 22, 18, 22, and 24 counts at $>$10~keV compared to an average
of 4 counts at $>$10~keV for the other 13 sources.  In addition, adding
the \cite{davis01} pile-up model to the spectral model significantly
improves the fits for the 5 piled-up spectra.

\begin{figure}
\includegraphics[clip,scale=0.45]{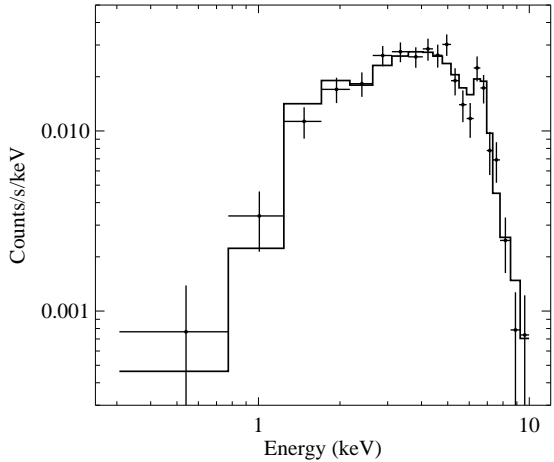}
\caption{{\em Chandra} spectrum for CXOU J143308.3--611539 (= IGR J14331--6112), which
has been confirmed as a High-Mass X-ray Binary based on optical spectroscopy.  The
X-ray spectrum is fitted with an absorbed power-law continuum and a Gaussian iron line
with an equivalent width of $\sim$945~eV.\label{fig:spectrum1}}
\end{figure}

Thus, after accounting for pile-up, 17 of the 18 sources are consistent
with a power-law model with various levels of absorption.  The one source
that is not consistent with a power-law is CXOU J143308.3--611539, for 
which $\chi^{2}_{\nu} = 2.5$ for 10 dof, and its residuals show a large 
excess in the energy bin corresponding to the Fe K$\alpha$ transition 
energy.  This likely indicates the presence of an Fe emission line, and
adding a Gaussian with an energy of $E_{\rm line} = 6.70^{+0.25}_{-0.17}$ keV, 
a width of $\sigma_{\rm line} = 0.27^{+0.21}_{-0.14}$ keV, and a flux of
$N_{\rm line} = (7.0^{+4.8}_{-3.4})\times 10^{-5}$ photons~cm$^{-2}$~s$^{-1}$ 
improves the quality of the fit to $\chi^{2}_{\nu} = 1.0$ for 7 dof.  
Figure~\ref{fig:spectrum1} shows the {\em Chandra} spectrum (rebinned 
to 21 bins rather than 13 bins) for CXOU J143308.3--611539, and the 
presence of the iron line, which has a very large equivalent width of 
945~eV, is clear.

\begin{table*}
\caption{{\em Chandra} Spectral Results\label{tab:spectra}}
\begin{minipage}{\linewidth}
\begin{center}
\begin{tabular}{cccccccc} \hline \hline
CXOU Name &
$N_{\rm H}$ ($\times 10^{22}$ cm$^{-2}$)\footnote{The parameters are for power-law fits to the {\em Chandra}/ACIS spectra and include photoelectric absorption with cross sections from \cite{bm92} and abundances from \cite{wam00}.  In general, the measured value of $N_{\rm H}$ and the errors on $N_{\rm H}$ are scaled-down by $\sim$30\% if solar abundances \citep{ag89} rather than the approximation to average interstellar abundances \citep{wam00}.  A pile-up correction was applied in the cases where pile-up parameters are given.  The PSF fraction \citep{davis01} was left fixed to 0.95 (the default value) except for the CXOU J194356.2+211823 spectrum, where it was left as a free parameter.  In this case, the PSF fraction dropped to $0.35^{+0.26}_{-0.13}$, presumably because this is a bright source that is relatively far (4.9 arcminutes) off-axis.  We performed fits without re-binning the data and using Cash statistics.  Errors in this table are at the 90\% confidence level ($\Delta$$C = 2.7$).} &
$\Gamma^{a}$ & 
X-ray Flux\footnote{Unabsorbed 0.3--10 keV flux in units of $10^{-12}$ erg~cm$^{-2}$~s$^{-1}$.} &  
$\alpha$\footnote{The grade migration parameter in the pile-up model \citep{davis01}. The probability that $n$ events will be piled together but will still be retained after data filtering is $\alpha^{n-1}$.} & 
Fit Statistic\footnote{The Cash statistic and degrees of freedom for the best fit model.} & 
Galactic $N_{\rm H}$/$N_{\rm H_{2}}$ ($\times 10^{22}$ cm$^{-2}$)\footnote{The atomic hydrogen column density through the Galaxy from \cite{kalberla05}.  We also give the molecular hydrogen column density through the Galaxy, using a CO map and conversion to $N_{\rm H_{2}}$ \citep{dht01}.} & 
$N_{\rm H,local}$ ($\times 10^{22}$ cm$^{-2}$)\footnote{The estimate for the column density due to material local to the source (see text for details on how the limits are derived).  An upper limit indicates no evidence for local absorption (although it also cannot be ruled out in any case), and a range indicates evidence for slight or significant local absorption.}\\ \hline \hline
J110926.4--650224 & $0.65^{+0.18}_{-0.17}$ & $1.43\pm 0.17$ & $4.8^{+0.4}_{-0.3}$ & -- & 530.6/660 & 0.52/0.085 & $<$0.83\\
J131825.0--625815 & $18^{+66}_{-13}$ & $1.9^{+6.3}_{-2.0}$ & $1.0^{+47.9}_{-0.6}$ & -- & 149.2/660 & 1.26/0.95 & 2--84\\
J143308.3--611539\footnote{A strong iron line is present and is included in the model as a Gaussian.  The parameters are $E_{\rm line} = 6.70^{+0.25}_{-0.17}$ keV, $\sigma_{\rm line} = 0.27^{+0.21}_{-0.14}$ keV, and $N_{\rm line} = (7.0^{+4.8}_{-3.4})\times 10^{-5}$ photons~cm$^{-2}$~s$^{-1}$, with an equivalent width of 945 eV.} & $2.2^{+0.9}_{-0.8}$ & $0.34^{+0.37}_{-0.33}$ & $6.3^{+0.7}_{-0.6}$ & -- & 614.2/657 & 1.51/0.53 & $<$3.1\\
J144628.2--641624 & $1.21^{+0.22}_{-0.40}$ & $1.50^{+0.20}_{-0.45}$ & $7.1^{+1.8}_{-0.6}$ & $1.0^{+0.0}_{-0.6}$ & 653.5/659 & 0.38/0.00 & 0.4--1.4\\
J155246.9--502953 & $0.59^{+0.41}_{-0.25}$ & $0.60^{+0.30}_{-0.28}$ & $1.8^{+0.3}_{-0.2}$ & -- & 522.0/660 & 0.68/0.050 & $<$1.0\\
J162826.8--502239 & $0.35^{+0.90}_{-0.35}$ & --$0.82^{+0.32}_{-0.26}$ & $6.8^{+0.9}_{-0.8}$ & -- & 637.0/660 & 1.37/0.45 & $<$1.3\\
J173518.7--325428 & $2.6^{+1.4}_{-1.2}$ & $0.79^{+0.56}_{-0.53}$ & $1.44^{+0.26}_{-0.22}$ & -- & 396.9/660 & 1.20/2.28 & $<$4.0\\
J173527.5--325554 & $7.5^{+3.0}_{-2.5}$ & $0.54^{+0.60}_{-0.55}$ & $13.1^{+2.5}_{-2.0}$ & $0.55^{+0.30}_{-0.40}$ & 541.0/659 & 1.18/2.28 & $<$11\\ 
J174026.8--365537 & $0.1^{+0.4}_{-0.1}$ & --$0.30^{+0.30}_{-0.24}$ & $7.2^{+1.3}_{-1.1}$ & -- & 507.7/660 & 0.42/0.11 & $<$0.5\\
J174437.3--323222\footnote{This is the spectrum of the point source a few arcminutes from the center of the nebula.} & $2.4^{+0.8}_{-0.6}$ & $0.97^{+0.34}_{-0.32}$ & $3.5\pm 0.4$ & -- & 526.0/660 & 0.67/0.47 & 0.2--3.2\\
J175039.4--264436 & $13.4^{+7.8}_{-5.5}$ & $0.44^{+0.84}_{-0.72}$ & $4.5^{+1.9}_{-0.7}$ & -- & 469.2/660 & 1.16/4.66 & $<$21\\
J175834.5--212321 & $15.6^{+6.0}_{-5.0}$ & $0.23^{+0.59}_{-0.54}$ & $11.4^{+2.3}_{-1.4}$ & -- & 516.0/660 & 0.76/0.48 & 9--22\\
J181328.0--163548 & $11.0^{+5.6}_{-4.1}$ & $1.44^{+0.89}_{-0.79}$ & $3.0^{+5.0}_{-2.1}$ & -- & 363.6/660 & 1.14/0.80 & 4--17\\
J181722.1--250842 & $0.11^{+0.35}_{-0.11}$ & --$0.28^{+0.19}_{-0.28}$ & $8.5^{+2.3}_{-1.1}$ & $1.0^{+0.0}_{-0.8}$ & 607.0/659 & 0.23/0.00 & $<$0.46\\
J183049.9--123219 & $0.40^{+0.29}_{-0.25}$ & $0.54^{+0.18}_{-0.31}$ & $9.5^{+3.2}_{-1.4}$ & $0.88^{+0.12}_{-0.48}$ & 588.7/659 & 1.02/0.66 & $<$0.69\\
J184825.4--004635 & $68^{+28}_{-23}$ & $3.0^{+1.9}_{-1.6}$ & $124^{+11,949}_{-116}$ & -- & 298.1/660 & 1.83/1.14 & 41--96\\
J192626.9+132205 & $0.31^{+0.13}_{-0.12}$ & $0.75\pm 0.12$ & $9.1^{+0.6}_{-0.5}$ & -- & 647.2/660 & 0.95/0.37 & $<$0.44\\
J194356.2+211823 & $1.89^{+0.25}_{-0.22}$ & $1.83\pm 0.11$ & $29^{+24}_{-5}$ & $0.68^{+0.32}_{-0.18}$ & 693.5/658 & 0.84/0.054 & 0.7--2.1\\ \hline
\end{tabular}
\end{center}
\end{minipage}
\end{table*}

\begin{figure*}
\begin{center}
\includegraphics[scale=0.9]{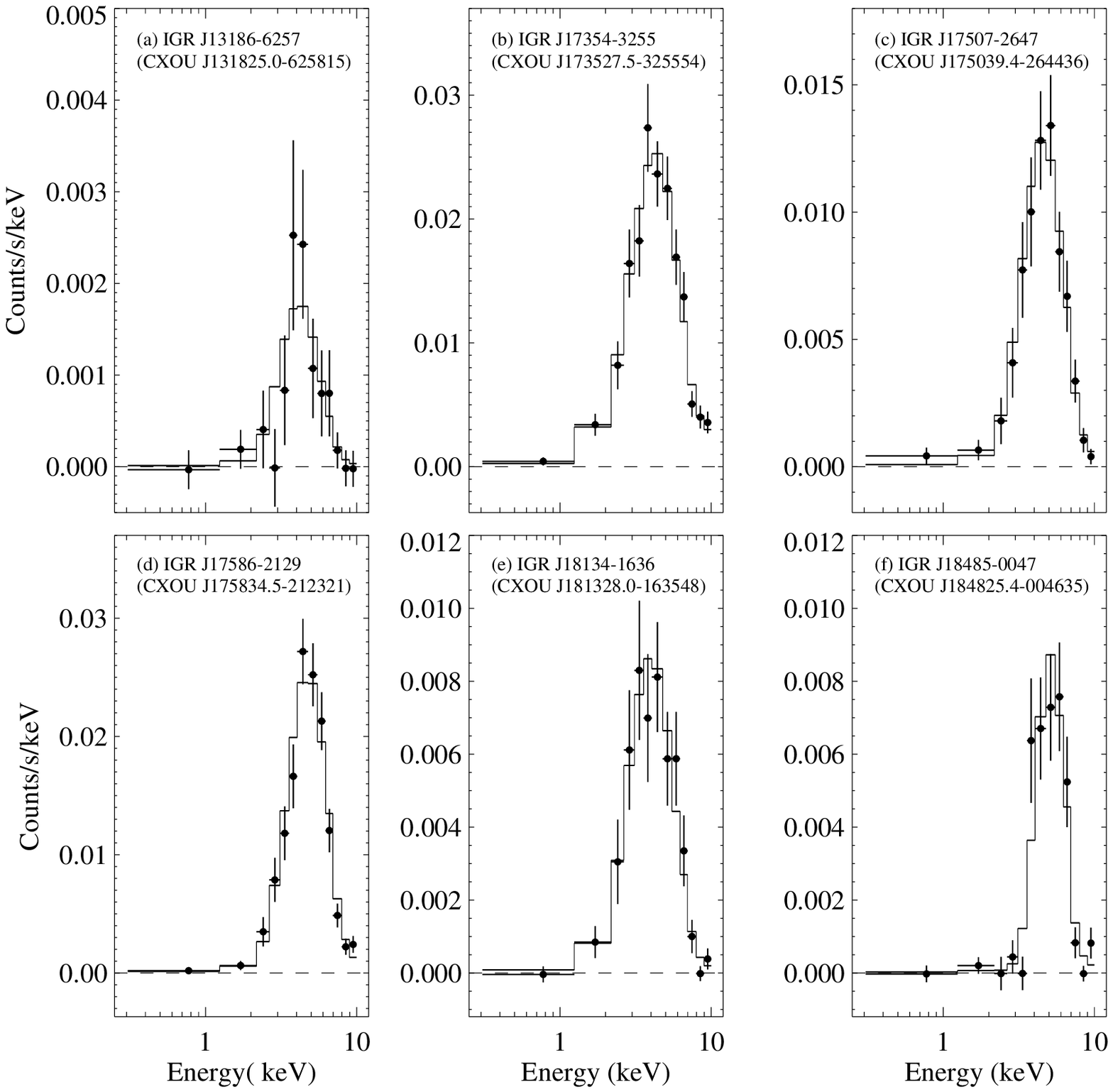}
\caption{{\em Chandra} ACIS-I spectra for the six sources that show the highest column density
measurements.  IGR~J13186--6257 {\it (a)}, IGR~J17586--2129 {\it (d)}, IGR~J18134--1636 
{\it (e)}, and IGR~J18485--0047 {\it (f)} require levels of $N_{\rm H}$ in excess of the 
Galactic levels (see Table~\ref{tab:spectra}) while IGR~J17354--3255 {\it (b)} and
IGR~J17507--2647 {\it (c)} may or may not be locally absorbed.\label{fig:spectra6}}
\end{center}
\end{figure*}

After determining the spectral models with the binned spectra, we refitted
all 18 of the spectra using Cash statistics \citep{cash79} and unbinned
spectra, which have 663 bins after restricting the energy range to 
0.3--10 keV.  The best fit parameters with 90\% confidence errors are 
given in Table~\ref{tab:spectra} along with the Galactic $N_{\rm H}$ and 
$N_{\rm H_{2}}$ values for each source \citep{kalberla05,dht01}.  The column 
densities measured by {\em Chandra} range from being significantly 
less than the Galactic values to being consistent with sources that are
extremely absorbed ($\sim$$10^{23-24}$ cm$^{-2}$).  To determine which spectra
show evidence for absorption by material local to the source, Table~\ref{tab:spectra}
includes limits on local absorption for each source.  In each case, the 
upper limit of this quantity, $N_{\rm H,local}$, is the upper end of the 
90\% confidence error region for the column density measured by {\em Chandra}, 
while the lower limit is given by the lower end of this error region 
minus the maximum absorption that could be interstellar 
($N_{\rm H} + 2N_{\rm H_{2}}$, where these are the Galactic values).  
Figure~\ref{fig:spectra6} shows the spectra for the six sources that show 
the highest measured values of $N_{\rm H}$.  The sources CXOU J131825.0--625815, 
CXOU J175834.5--212321, CXOU J181328.0--163548, and CXOU J184825.4--004635
require local absorption at the 90\% confidence level with $N_{\rm H,local}$ 
ranges of (2--84)$\times 10^{22}$ cm$^{-2}$, (9--22)$\times 10^{22}$ cm$^{-2}$, 
(4--17)$\times 10^{22}$ cm$^{-2}$, and (41--96)$\times 10^{22}$ cm$^{-2}$.  The 
other two sources, CXOU J173527.5--325554 and CXOU J175039.4--264436 may or 
may not have significant levels of local absorption with upper limits on 
$N_{\rm H,local}$ of $<$$1.1\times 10^{23}$ cm$^{-2}$ and 
$<$$2.1\times 10^{23}$ cm$^{-2}$, respectively.

\subsection{Using X-ray Source Number Densities to Assess the Probability of Spurious Associations}

With the spectral parameters for the point sources determined, we can now 
calculate the probability that the candidate associations between the 
{\em Chandra} and IGR sources are spurious.  This calculation requires
knowledge of the number density of X-ray sources with fluxes between
$\sim$$10^{-13}$ and $\sim$$10^{-11}$ ergs~cm$^{-2}$~s$^{-1}$ in the Galactic 
plane.  The broadest survey to date that resulted in a suitable 
$\log{N}$-$\log{S}$ curve was done by the {\em Advanced Satellite for 
Cosmology and Astrophysics (ASCA)}.  The survey covers Galactic longitudes
from --$45^{\circ}$ to +$45^{\circ}$ and Galactic latitudes from --$0.4^{\circ}$
to +$0.4^{\circ}$.  The 2--10 keV $\log{N}$-$\log{S}$ is described as a 
power-law with $N(>F_{2-10~\rm keV}) = 9.2(F_{2-10~\rm keV}/10^{-13})^{-0.79}$ deg$^{-2}$, 
where $F_{2-10~\rm keV}$ is the absorbed 2--10~keV flux in units of ergs~cm$^{-2}$~s$^{-1}$
\citep{sugizaki01}.  The flux range for the {\em ASCA} survey is 
$2\times 10^{-13}$ to $10^{-10}$ ergs~cm$^{-2}$~s$^{-1}$ (2--10~keV).  
Smaller surveys of parts of the Galactic plane by {\em XMM-Newton} and 
{\em Chandra} that extend the $\log{N}$-$\log{S}$ to lower flux levels 
give $\log{N}$-$\log{S}$ curves that are consistent with {\em ASCA} in
the flux ranges where they overlap \citep{hands04,ebisawa05}.  

To calculate the probabilities of spurious associations, we first used
the spectral parameters for the 18 {\em Chandra} point source (see
Table~\ref{tab:spectra}) to determine the absorbed 2--10~keV flux for
each source, and these are reported in Table~\ref{tab:spurious}.  Then,
we used the source density expression given above to calculate 
$N(>F_{2-10~\rm keV})$ for each source.  To determine the area in which 
we searched to find the {\em Chandra} source, we consider both the
angular distance of the {\em Chandra} source from the center of the
{\em INTEGRAL} error circle ($\theta$ in Table~\ref{tab:spurious}) 
and the radius of the 90\% confidence {\em INTEGRAL} error circle
($\theta_{INTEGRAL}$ in Table~\ref{tab:spurious}), and we used the larger
of these two quantities to calculate the probability.  

As shown in Table~\ref{tab:spurious}, this analysis indicates a probability 
of a spurious association of $<$1\% in 14 cases and in the 1--2\% range
in 3 cases.  The faintest of the sources, CXOU J131825.0--625815, is the 
only one with an appreciable probability of a spurious association at 5.4\%.  
While this could indicate that the association is spurious, the calculation 
does not account for the fact that the source is a hard X-ray source and also 
that it is relatively close to the center of the {\em INTEGRAL} error circle.  
One other case to consider is that of IGR~J17354--3255, for which we found 
two {\em Chandra} sources within the {\em INTEGRAL} error circle.  Although 
it is possible that both sources contribute to the flux detected by 
{\em INTEGRAL}, it is still probably most reasonable to consider that one
of these is a spurious association.  While CXOU J131825.0--625815, 
CXOU J173518.7--325428, CXOU J173527.5--325554 would be the leading candidates 
for being spurious associations, the results of these calculations increase
our confidence in the associations between the {\em Chandra} and IGR sources.

\begin{table*}
\caption{Quantities for Calculating the Probability of Spurious Associations\label{tab:spurious}}
\begin{minipage}{\linewidth}
\begin{center}
\begin{tabular}{cccccc} \hline \hline
CXOU Name & 
$F_{2-10~\rm keV}$\footnote{The absorbed 2--10 keV flux in units of ergs~cm$^{-2}$~s$^{-1}$.} &
$N(>F_{2-10~\rm keV})$ (deg$^{-2}$)\footnote{The X-ray source density at the 2--10 keV flux level for each source from the {\em ASCA} survey of the Galactic plane \citep{sugizaki01}.} & 
$\theta$ (arcmin)\footnote{The angular distance between the center of the {\em INTEGRAL} error circle and the source.} & 
$\theta_{INTEGRAL}$ (arcmin)\footnote{The 90\% confidence {\em INTEGRAL} error radius \citep{bird06}.} &
Probability (\%)\\ \hline
J110926.4--650224 & $3.3\times 10^{-12}$ & 0.59 & 6.02 & 5.5 & 1.9\\
J131825.0--625815 & $2.6\times 10^{-13}$ & 4.32 & 1.94 & 3.8 & 5.4\\
J143308.3--611539 & $6.3\times 10^{-12}$ & 0.35 & 4.02 & 3.9 & 0.49\\
J144628.2--641624 & $4.5\times 10^{-12}$ & 0.46 & 1.46 & 3.7 & 0.55\\
J155246.9--502953 & $1.6\times 10^{-12}$ & 1.05 & 1.59 & 3.9 & 1.4\\
J162826.8--502239 & $6.7\times 10^{-12}$ & 0.33 & 3.12 & 4.4 & 0.56\\
J173518.7--325428 & $1.1\times 10^{-12}$ & 1.35 & 1.83 & 2.2 & 0.57\\
J173527.5--325554 & $9.7\times 10^{-12}$ & 0.25 & 1.34 & 2.2 & 0.11\\
J174026.8--365537 & $7.0\times 10^{-12}$ & 0.32 & 0.84 & 3.5 & 0.34\\
J174437.3--323222 & $2.7\times 10^{-12}$ & 0.68 & 3.77 & 2.2 & 0.84\\
J175039.4--264436 & $3.0\times 10^{-12}$ & 0.62 & 2.97 & 2.6 & 0.48\\
J175834.5--212321 & $7.7\times 10^{-12}$ & 0.30 & 3.81 & 3.3 & 0.38\\
J181328.0--163548 & $1.4\times 10^{-12}$ & 1.16 & 0.97 & 3.8 & 1.5\\
J181722.1--250842 & $8.4\times 10^{-12}$ & 0.28 & 0.86 & 2.2 & 0.12\\
J183049.9--123219 & $8.6\times 10^{-12}$ & 0.27 & 0.83 & 3.3 & 0.26\\
J184825.4--004635 & $1.8\times 10^{-12}$ & 0.96 & 0.55 & 3.4 & 0.97\\
J192626.9+132205 &  $7.9\times 10^{-12}$ & 0.29 & 4.79 & 3.7 & 0.58\\
J194356.2+211823 &  $1.4\times 10^{-11}$ & 0.19 & 4.88 & 4.9 & 0.39\\ \hline
\end{tabular}
\end{center}
\end{minipage}
\end{table*}

\vspace{3cm}
\section{Discussion}

\subsection{Extended Sources}

\subsubsection{IGR J14003--6326}

\begin{figure*}
\begin{center}
\includegraphics[clip, scale=0.9]{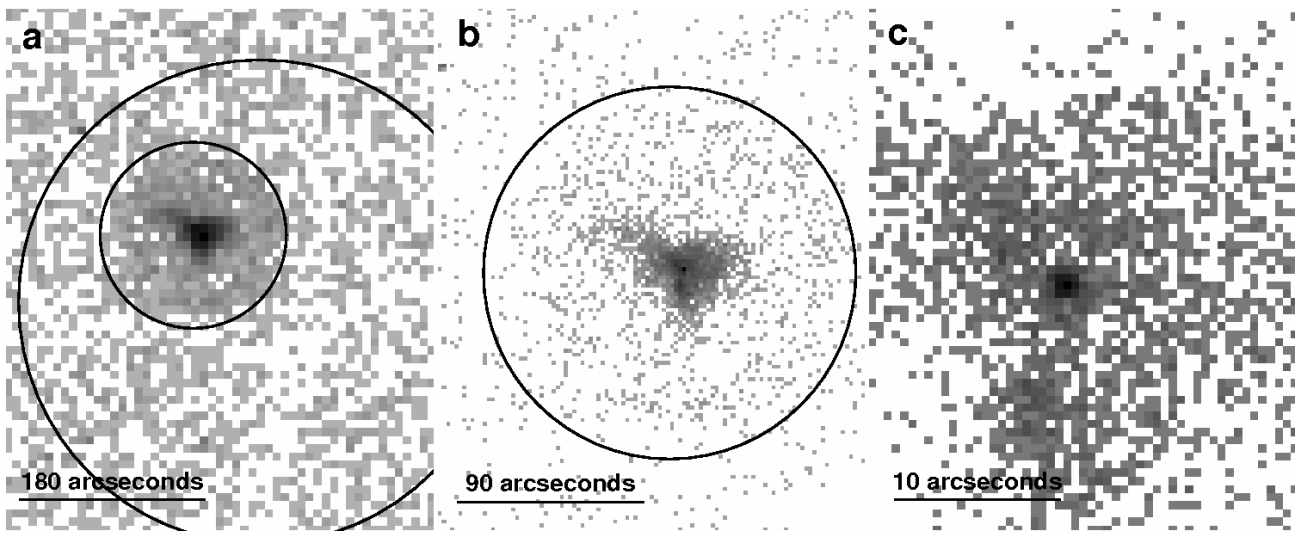}
\caption{{\em Chandra} 0.3--10 keV images for IGR J14003--6326.  {\it (a)} The pixel size in this image is 
$7^{\prime\prime}.9$.  The larger circle shows the 90\% confidence {\em INTEGRAL} error circle, which has a
radius of $3^{\prime}.9$.  The smaller circle has a radius of $1^{\prime}.5$, and marks the approximate extent
of the supernova remnant (SNR).  {\it (b)} The pixel size in this image is $1^{\prime\prime}.97$, and the image 
is meant to highlight the putative pulsar wind nebula.  The $1^{\prime}.5$ radius circle is shown. {\it (c)}
The pixel size in this unbinned image is $0^{\prime\prime}.492$, and the image is meant to highlight the 
point-like source near the center of the SNR.  In all 3 images, North is up and East is to the left.
\label{fig:3panels}}
\end{center}
\end{figure*}

IGR J14003--6326 was discovered by {\em INTEGRAL} in 2006 \citep{keek06}.  
Although it was observed by {\em Swift}, no conclusion was reached on its 
nature \citep{malizia07}.  Figure~\ref{fig:3panels} shows 0.3--10 keV 
{\em Chandra}/ACIS images of the IGR J14003--6326 field.  The widest view 
(Figure~\ref{fig:3panels}a) shows the presence of a nearly circular source 
with a radius of $1^{\prime}.5$, and this is the only bright source detected 
in the $3^{\prime}.9$ radius {\em INTEGRAL} error circle.  The position and
the fact that it is a hard source make it essentially certain that this is 
the IGR source, and the morphology and hardness allow us to identify this 
source as a supernova remnant (SNR).  For Figure~\ref{fig:3panels}b, we 
zoom-in toward the center of the SNR to show that there is significant 
structure near the center at size scales of tens of arcseconds.  We suspect 
that this is a pulsar wind nebula (PWN).  Finally, Figure~\ref{fig:3panels}c 
shows the {\em Chandra} image without any pixel binning, and it is clear that 
there is a point-like source near the center of the SNR where a putative 
neutron star might reside.  The number of counts in the brightest pixel is 82, 
and the central 9 pixels have 298 counts.  The position of this point-like
source is R.A.~(J2000) = $14^{\rm h}00^{\rm m}45^{\rm s}.69$, 
Decl.~(J2000) = --$63^{\circ}25^{\prime}42^{\prime\prime}\!.6$, and it is very
close to the center of the SNR.  Despite the relative brightness of the 
point-like source, this is only a small fraction of the 4075 counts
collected for the entire SNR.  Thus, the emission detected by {\em INTEGRAL}
likely comes from the PWN or the SNR.

We have used the CIAO routine {\ttfamily specextract} to produce a spectrum
from a circular region centered on the point-like source with a radius of
$1^{\prime}.5$, while the background is taken from a source-free rectangular
region on the same ACIS-I chip.  The 0.3--10 keV energy spectrum is 
well-described by an absorbed power-law, with this model giving 
$\chi^{2}_{\nu} = 0.92$ for 161 dof.  The measured $N_{\rm H}$ is 
$(3.1\pm 0.3)\times 10^{22}$ cm$^{-2}$, which is somewhat higher than 
the Galactic value ($N_{\rm H} + 2N_{\rm H_{2}}$) of $2.1\times 10^{22}$ cm$^{-2}$. 
The measured power-law photon index is $\Gamma = 1.83\pm 0.13$, and the 
unabsorbed 0.3--10 keV flux is $3.6\times 10^{-11}$ ergs~cm$^{-2}$~s$^{-1}$.  
The unabsorbed 2--10 keV flux is $1.9\times 10^{-11}$ ergs~cm$^{-2}$~s$^{-1}$,
which corresponds to $\sim$1 millicrab.  

There has only been one other IGR source identified as a SNR, IGR~J18135--1751, 
and it has been identified as the TeV source HESS~J1813-178 \citep{ubertini05}.
Searches of the {\em High Energy Stereoscopic System (HESS)} source 
catalog\footnote{See http://www.mpi-hd.mpg.de/hfm/HESS/pages/home/sources.}
as well as the SIMBAD database and the {\em Fermi} Large Area Bright Gamma-Ray
Source List \citep{abdo09} do not indicate that IGR~J14003--6326 is identified 
with any TeV or GeV source.  The 20--100~keV flux for IGR~J14003--6326 is 
$1.8\times 10^{-11}$ ergs~cm$^{-2}$~s$^{-1}$ \citep{bird06}, which is only slightly 
less than the $3\times 10^{-11}$ ergs~cm$^{-2}$~s$^{-1}$ flux of IGR~J18135--1751 
\citep{ubertini05}.  However, \cite{ubertini05} show that the TeV emission for 
IGR~J18135--1751 is more than an order of magnitude in excess of an extrapolation 
of the X-ray power-law, and \cite{helfand07} point out that the TeV-to-X-ray
luminosity ratio for IGR~J18135--1751 is extreme with a value close to unity, 
which is much higher than the ratio of 0.06 measured for the Crab Nebula and
also significantly higher than the highest ratios known for PWNe (e.g., 
0.5 for PSR 1509--58 and 0.3 for Vela X).  It has been suggested that the 
possible proximity to the star forming region W33 to IGR~J18135--1751 may 
provide a higher density of seed photons for Inverse Compton scattering to
TeV energies \citep{funk07,helfand07}.  Thus, the non-detection of TeV 
emission from IGR~J14003--6326 is not surprising and suggests that its high 
energy properties are more similar to the previously known PWN than to 
IGR~J18135--1751.

\subsubsection{IGR J17448--3232}

\begin{figure*}
\begin{center}
\includegraphics[clip, scale=0.9]{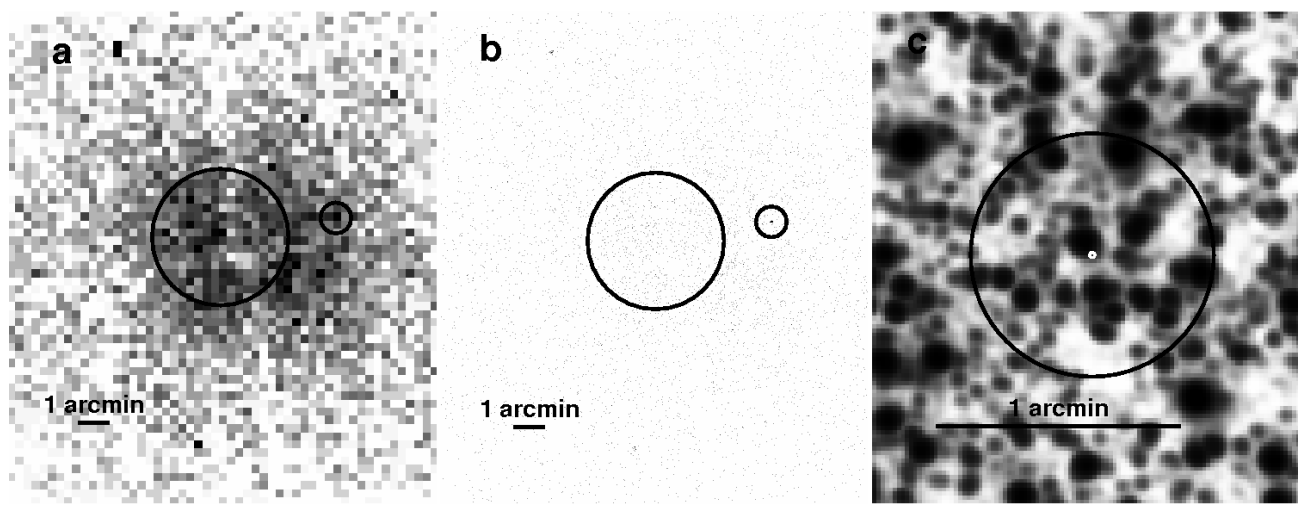}
\caption{Images of the IGR J17448--3232 field.  {\it (a)} A {\em Chandra} 0.3--10 keV image rebinned to
a pixel size of $15^{\prime\prime}.7$.  The larger circle shows the 90\% confidence {\em INTEGRAL} error 
circle, which has a radius of $2^{\prime}.2$.  The smaller circle has a radius of $0^{\prime}.5$, and marks 
the location of the hard point source, CXOU J174437.3--323222.  The extended source is likely a supernova 
remnant (SNR), and we suggest that the hard point source may be an associated pulsar.  {\it (b)} The 
pixel size in this image is $0^{\prime\prime}.98$.  The same two circles are plotted, and the image is meant 
to highlight the hard point source.  {\it (c)} A 2MASS $K_{s}$-band image of the region near the hard point
source.  The black circle is the same $0^{\prime}.5$-radius circle shown in the other two panels, and the 
{\em Chandra} 90\% confidence error circle is shown in white.  In all 3 images, North is up and East is 
to the left.\label{fig:j17448}}
\end{center}
\end{figure*}

From a {\em Swift} observation, \cite{landi1323} mentioned the presence of
diffuse emission in the IGR J17448--3232 field, and our {\em Chandra} 
observation confirms this.  Figure~\ref{fig:j17448}a shows a rebinned
0.3--10 keV image along with the $2^{\prime}.2$ radius {\em INTEGRAL} error
circle, which is centered on the extended source.  Like the previous source, 
the extended source associated with IGR J17448--3232 is a nearly circular, 
and is likely a SNR.  However, we do not see any evidence for a PWN within
the SNR, and the only point-like source is CXOU J174437.3--323222, which is
a hard source at the edge of the SNR (see Figure~\ref{fig:j17448}b).  We
extracted a {\em Chandra} spectrum from the extended source, using a circular
extraction region with a radius of $3^{\prime}$.  The spectrum is well-described
($\chi^{2}_{\nu} = 0.99$ for 163 dof) by an absorbed power-law with 
$N_{\rm H} = (2.5^{+0.5}_{-0.4})\times 10^{22}$ cm$^{-2}$, which is somewhat above 
the interstellar column density ($N_{\rm H} + 2N_{\rm H_{2}}$) of 
$1.6\times 10^{22}$ cm$^{-2}$.  The power-law photon index is 
$\Gamma = 1.27^{+0.22}_{-0.21}$, and the 0.3--10 keV unabsorbed flux is
$2.3\times 10^{-11}$ ergs~cm$^{-2}$~s$^{-1}$.  We do not find any high-energy
counterparts to this SNR in the {\em HESS} or {\em Fermi} catalogs.
As discussed above, even though TeV emission is detected for 
IGR~J18135--1751, the non-detection of TeV emission for IGR~J17448--3232
is not surprising because IGR~J18135--1751 is an extreme case.  In addition,
IGR~J17448--3232 may be even less likely to emit strong TeV emission because
we do not see a PWN within the SNR.

The fact that the position at the center of the extended source agrees well with 
the {\em INTEGRAL} position argues that the SNR is primarily producing the 
$1.1\times 10^{-11}$ ergs~cm$^{-2}$~s$^{-1}$ flux that {\em INTEGRAL} is detecting 
in the 20--100~keV band \citep{bird06}.  However, it is still possible that the 
hard point source, CXOU J174437.3--323222, is associated with the SNR.  The hard 
point source does not have counterparts in the 2MASS, DENIS, USNO-B1.0, or USNO-A2.0 
catalogs; however, the $K_{s}$-band image shown in Figure~\ref{fig:j17448}c shows that 
the {\em Chandra} position is on the edge of the point spread function of a relatively 
bright IR source, so better optical/IR images are necessary.  If CXOU J174437.3--323222 
is an isolated neutron star that received a kick when the supernova occurred, it 
should be very faint in the optical and IR.  While the association between the SNR 
and CXOU J174437.3--323222 is intriguing, further work is necessary for confirmation.

\subsection{Point Sources}

\begin{figure*}
\begin{center}
\includegraphics[clip, scale=0.9]{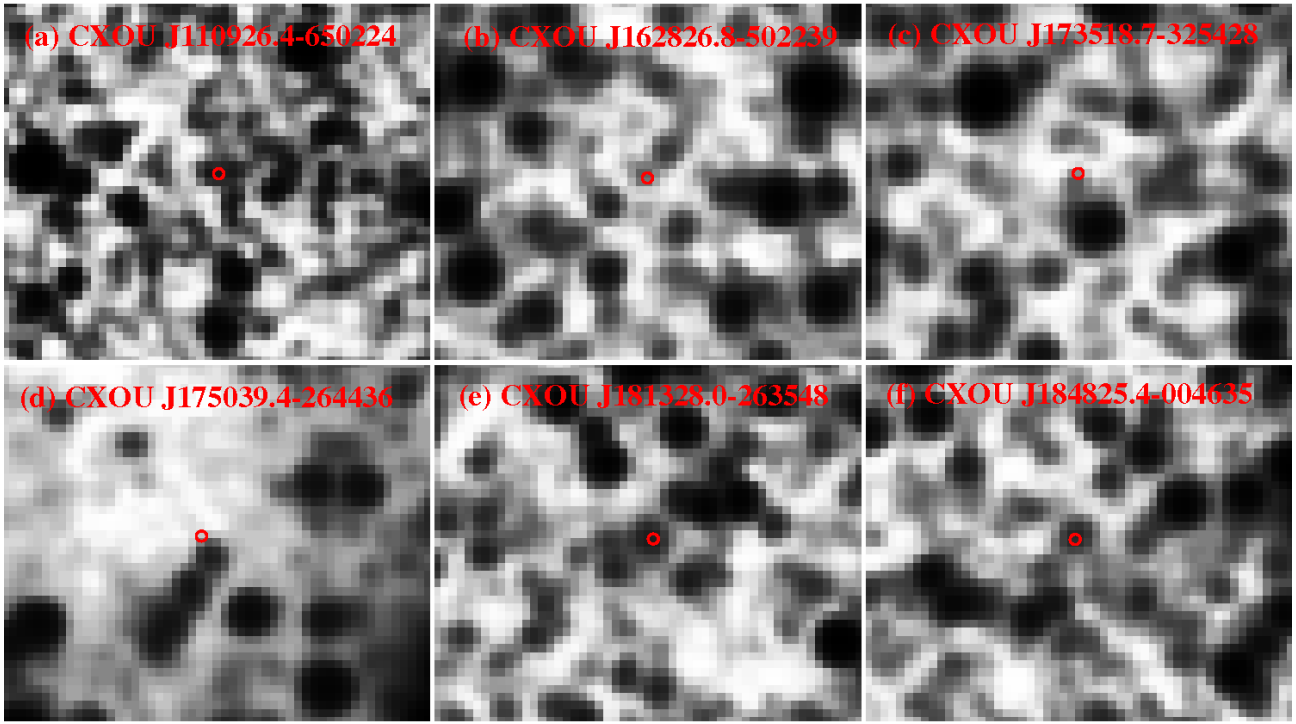}
\caption{2MASS $K_{s}$-band images with {\em Chandra} 90\% confidence error circles for 6 IGR 
sources for which there are no counterparts in the 2MASS, DENIS, USNO-B1.0, or USNO-A2.0 catalogs.
However, some of the images, {\it (a)}, {\it (b)}, {\it (e)}, and {\it (f)} suggest that there 
are faint IR counterparts.  The images are for the following sources:
{\it (a)} CXOU J110926.4--650224, {\it (b)} CXOU J162826.8--502239, {\it (c)} CXOU J173518.7--325428, 
{\it (d)} CXOU J175039.4--264436, {\it (e)} CXOU J181328.0--263548, and {\it (f)} CXOU J184825.4--004635.
In each image, the pixel size is $1^{\prime\prime}$ and North is up and 
East is to the left.  The images are $52^{\prime\prime}$ in the East-West direction and 
$43^{\prime\prime}$ in the North-South direction.\label{fig:k06}}
\end{center}
\end{figure*}

{\bf IGR J11098--6457:}  Based on an earlier {\em Swift} observation, two 
possible counterparts were considered, and it was found that 
1RXS J110927.4--650245 is the more likely counterpart due to its hard
X-ray spectrum \citep{landi1538}.  The position of CXOU J110926.4--650224, 
the source that we find to be the most likely counterpart, is consistent 
with the {\em Swift} position.  There are no 2MASS, DENIS, USNO-B1.0, 
or USNO-A2.0 sources consistent with the {\em Chandra} position.  However, 
the 2MASS $K_{s}$-band image shown in Figure~\ref{fig:k06}a shows that 
there appears to be a faint IR source at the {\em Chandra} position.  The
{\em Chandra} energy spectrum is consistent with a power-law with a photon
index of $\Gamma = 1.43\pm 0.17$ and a column density consistent with the
Galactic column density (see Table~\ref{tab:spectra}).  The 2--10 keV
absorbed flux is $3.3\times 10^{-12}$ ergs~cm$^{-2}$~s$^{-1}$.  Thus, the
spectrum measured by {\em Chandra} is consistent with that measured by
{\em Swift}, which found $\Gamma\sim 1.2$ and a 2--10 keV flux of
$2.7\times 10^{-12}$ ergs~cm$^{-2}$~s$^{-1}$ \citep{landi1538}.

\begin{figure*}
\begin{center}
\includegraphics[clip, scale=0.9]{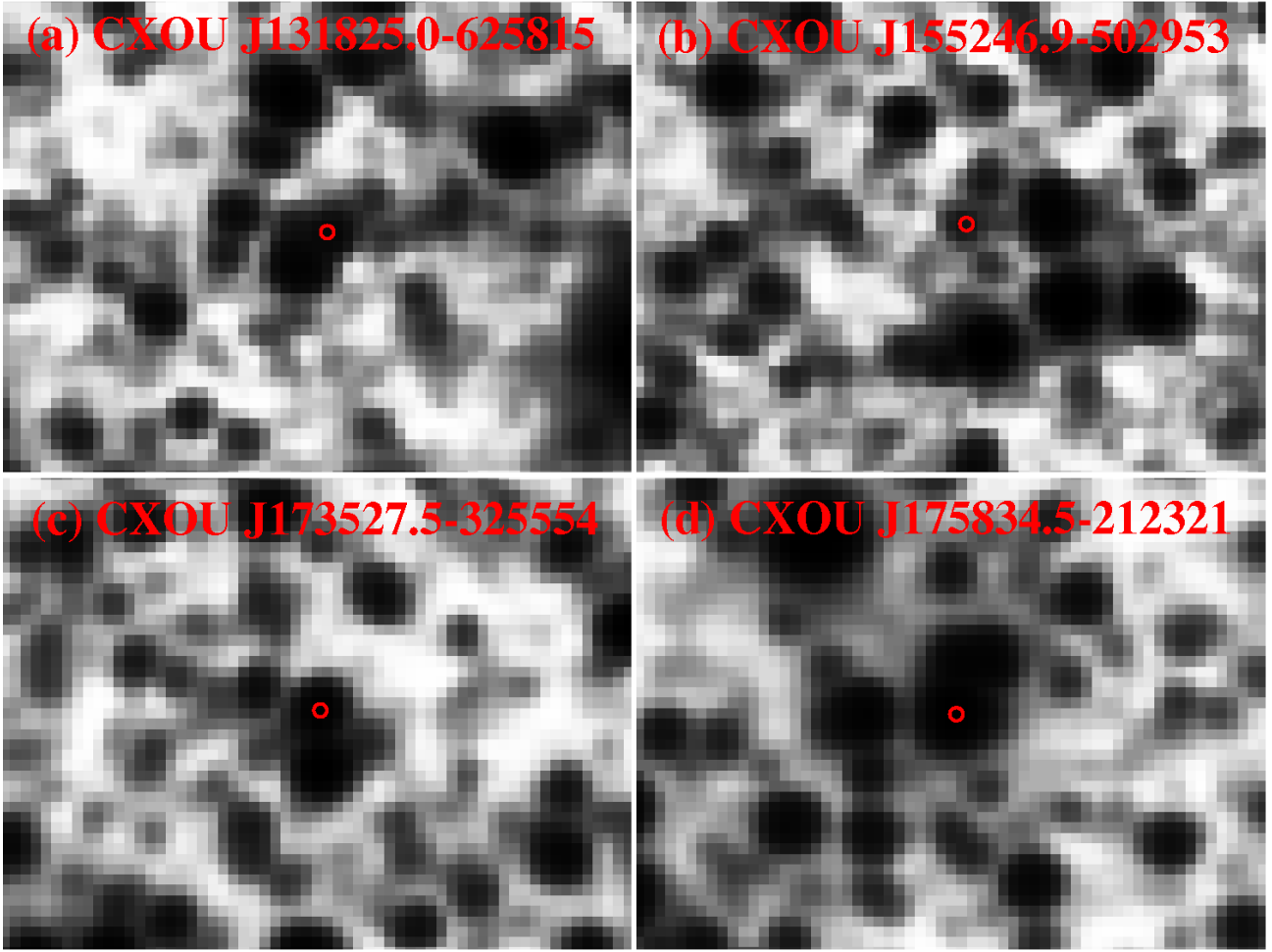}
\caption{2MASS $K_{s}$-band images with {\em Chandra} 90\% confidence error circles for the 4 IGR 
sources with new optical/IR identifications (see Table~\ref{tab:oir}).  The images are for the
following sources:  {\it (a)} CXOU J131825.0--625815, {\it (b)} CXOU J155246.9--502953, 
{\it (c)} CXOU J173527.5--325554, and {\it (d)} CXOU J175834.5--212321.  In each image, the pixel 
size is $1^{\prime\prime}$ and North is up and East is to the left.  The images are $60^{\prime\prime}$ 
in the East-West direction and $45^{\prime\prime}$ in the North-South direction.\label{fig:k04}}
\end{center}
\end{figure*}

{\bf IGR J13186--6257:}  The position of CXOU J131825.0--625815 is consistent
with the {\em Swift} position for this IGR source reported in \cite{landi1539}.
The refined {\em Chandra} position allows us to identify the IR counterpart as 
2MASS J13182505--6258156, which is $0^{\prime\prime}.2$ from the {\em Chandra} 
position and has IR magnitudes of $J = 13.58$, $H = 12.69$, 
$K_{s} = 12.84\pm 0.05$ (see Table~\ref{tab:oir} and Figure~\ref{fig:k04}a).  
There are no corresponding sources in the DENIS, USNO-B1.0, or USNO-A2.0 catalogs.  
Although \cite{landi1539} report that the {\em Swift} spectrum of this source is 
unabsorbed, we measure $N_{\rm H} = (1.8^{+6.6}_{-1.3})\times 10^{23}$ cm$^{-2}$, and
we find that $N_{\rm H,local}$ is in the range (2--84)$\times 10^{22}$ cm$^{-2}$
(see Table~\ref{tab:spectra}).  While we cannot make a definitive statement about 
the nature of the source, high and variable column densities have been seen for 
several of the IGR HMXBs.

{\bf IGR J14331--6112:}  A {\em Swift} position was previously obtained for
this source \citep{landi1273}, leading to a likely identification with a
USNO-B1.0 source.  Follow-up optical spectroscopy indicated a HMXB harboring
an optical companion with a B~III or B~V spectral type \citep{masetti6}.  
The position of CXOU J143308.3--611539 is consistent with the HMXB, confirming
the identification.  The {\em Chandra} energy spectrum is consistent with a 
power-law with a photon index of $\Gamma = 0.34^{+0.37}_{-0.33}$, which is
similar to the value of $\Gamma\sim 0.7$ quoted for the {\em Swift} spectrum.
However, a new result from the {\em Chandra} spectrum is the discovery of
a very strong iron K$\alpha$ emission line with an equivalent width of
$\sim$945 eV.  Strong iron lines have been seen for some of the other IGR 
HMXBs \citep{mg03,walter03,walter06}, but they are usually associated
with the class of obscured HMXBs.  Although we measure a relatively low
column density of $N_{\rm H} = (2.2^{+0.9}_{-0.8})\times 10^{22}$ cm$^{-2}$, 
consistent with the Galactic value, for CXOU J143308.3--611539, it should
be noted that the presence of a soft excess can cause artificially low
column densities to be measured in spectra with lower statistical 
quality \citep{tomsick09}.  For the {\em Chandra} spectrum, if we add
a second power-law component to account for a putative soft excess 
\citep[as in][]{tomsick09}, the find that the column density for the 
primary power-law component can be as high as $1.1\times 10^{23}$ cm$^{-2}$.

{\bf IGR J14471--6414:}  A {\em Swift} position was previously obtained for 
this source \citep{landi1273}, and follow-up optical spectroscopy indicated
the presence of a Seyfert 1.2 AGN with a redshift of $z = 0.053$ in the 
{\em Swift} error circle \citep{masetti6}.  The position of 
CXOU J144628.2--641624 is consistent with the position of the AGN, confirming
the association.

{\bf IGR J15529--5029:}  The position of CXOU J155246.9--502953 is consistent
with the {\em Swift} position for this IGR source reported in \cite{landi1539}.
The refined {\em Chandra} position allows us to identify the IR 
counterpart as 2MASS J15524694--5029534, which is $0^{\prime\prime}.2$ from the
{\em Chandra} position and has IR magnitudes of $J = 15.84\pm 0.10$, 
$H = 15.27\pm 0.13$, and $K_{s} = 14.77\pm 0.12$.  In addition, 
USNO-B1.0 0395-0509024 and USNO-A2.0 0375-25891641 are $0^{\prime\prime}.3$ 
from the {\em Chandra} position, and the optical magnitudes are listed as
$R = 18.6$ and $B = 21.4$ for the former and $R = 17.6$ and $B = 20.3$ for
the latter (see Table~\ref{tab:oir}).  Although we cannot conclude definitively 
on the nature of this source, the X-ray spectrum measured by {\em Swift} (a 
power-law with $\Gamma\sim 0.9$) and {\em Chandra} (a power-law with 
$\Gamma = 0.60^{+0.30}_{-0.28}$) are harder than would be expected for an AGN 
\citep{malizia07}, and we suggest that a CV or X-ray binary nature is more likely.

{\bf IGR J16287--5021:}  The {\em Chandra} information for CXOU J162826.8--502239
was reported previously in \cite{tomsick1649}, and the {\em Chandra} and {\em Swift} 
results for this source are discussed in detail in \cite{rtc09}.  Both 
{\em Chandra} and {\em Swift} detect the source at about the same flux level and 
with a low level of absorption.  However, the {\em Chandra} power-law index of
$\Gamma$ = --0.82$^{+0.32}_{-0.26}$ is significantly harder than the value of 
$\Gamma = 0.9\pm 0.8$ found by {\em Swift}.  There are no 2MASS, DENIS, USNO-B1.0, 
or USNO-A2.0 sources consistent with the {\em Chandra} position.  However, 
Figure~\ref{fig:k06}b shows the 2MASS $K_{s}$-band image for this field, and
there is a faint IR source that appears to be consistent with the {\em Chandra}
position.  The very hard X-ray spectrum suggests a CV or X-ray binary nature rather 
than an AGN nature, and perhaps the lack of an optical counterpart along with the 
faint IR candidate counterpart favors a relatively distant X-ray binary (likely an 
HMXB) rather than a CV.

{\bf IGR J17354--3255:}  This source was discovered within $\sim$$5^{\circ}$ of the 
Galactic center by {\em INTEGRAL} in 2006 \citep{kuulkers874,bird06}.  There are two 
relatively bright and hard {\em Chandra} sources within the {\em INTEGRAL} error 
circle \citep[see also][]{tomsick2022}, and these were also detected during a 
{\em Swift} observation made in 2009 April \citep{vercellone2019}.  The 
position of CXOU J173527.5--325554, the brighter and harder of the two sources, is 
within $0^{\prime\prime}.2$ of 2MASS J17352760--3255544, which has $J = 12.51\pm 0.05$, 
$H = 10.99\pm 0.04$, and $K_{s} = 10.27\pm 0.03$.  The source also appears in the 
DENIS catalog with similar $J$ and $K_{s}$ magnitudes (see Table~\ref{tab:oir}).  
However, it must be quite faint optically since it is not detected in the DENIS 
$I$-band, and it does not appear in the USNO catalogs.  The large $J-K_{s}$ value, 
high inferred optical extinction, and the relatively high X-ray column density 
$(7.5^{+3.0}_{-2.5})\times 10^{22}$ cm$^{-2}$ argue that the source is distant, probably 
either an X-ray binary near the Galactic center or an AGN.  However, as argued above, 
the hard power-law index of $\Gamma = 0.54^{+0.60}_{-0.55}$ argues against the AGN 
possibility and for the possibility that the source is an HMXB.  The other 
{\em Chandra} source, CXOU J173518.7--325428, does not have 2MASS, DENIS, USNO-B1.0, 
or USNO-A2.0 counterparts.  Although the $K_{s}$-band image (see Figure~\ref{fig:k06}c) 
shows a faint IR source within a couple arcseconds, it is not close enough to be 
associated with the {\em Chandra} source.  It is possible that both {\em Chandra} 
sources contribute to the flux detected by {\em INTEGRAL}, but CXOU J173527.5--325554 
would probably be a more interesting source for any further follow-up work.  Another
reason for interest in pursuing the nature of IGR~J17354--3255 is that its position
is consistent with the position of the gamma-ray transient AGL J1734--3310
\citep{bulgarelli2017}.

\begin{table*}
\caption{New Optical/Infrared Identifications\label{tab:oir}}
\begin{minipage}{\linewidth}
\begin{center}
\begin{tabular}{lcccc} \hline \hline
Catalog/Source\footnote{The catalogs are the 2 Micron All-Sky Survey (2MASS), the Deep Near Infrared Survey of the Southern Sky (DENIS), and the United States Naval Observatory (USNO-B1.0 and USNO-A2.0). The 2MASS $K_{s}$-band images for these four sources are shown in Figure~\ref{fig:k04}} & 
Separation & 
& 
Magnitudes & \\ \hline
IGR J13186--6257/CXOU J131825.0--625815 & & & & \\ \hline
2MASS J13182505--6258156 & $0^{\prime\prime}.2$ & $J = 13.58$ & $H = 12.69$ & $K_{s} = 12.84\pm 0.05$\\ \hline
IGR J15529--5029/CXOU J155246.9--502953 & & & & \\ \hline
2MASS J15524694--5029534 & $0^{\prime\prime}.2$ & $J = 15.84\pm 0.10$ & $H = 15.27\pm 0.13$ & $K_{s} = 14.77\pm 0.12$\\
USNO-B1.0 0395-0509024   & $0^{\prime\prime}.3$ & $B = 21.4\pm 0.3$ & $R = 18.6\pm 0.3$ & ---\\
USNO-A2.0 0375-25891641  & $0^{\prime\prime}.3$ & $B = 20.3\pm 0.5$ & $R = 17.6\pm 0.4$ & ---\\ \hline
IGR J17354--3255/CXOU J173527.5--325554 & & & & \\ \hline
2MASS J17352760--3255544 & $0^{\prime\prime}.2$ & $J = 12.51\pm 0.05$ & $H = 10.99\pm 0.04$ & $K_{s} = 10.27\pm 0.03$\\
DENIS J173527.6--325554  & $0^{\prime\prime}.2$ & $J = 12.36\pm 0.07$ & ---                 & $K_{s} = 10.12\pm 0.06$\\ \hline
IGR J17586--2129/CXOU J175834.5--212321 & & & & \\ \hline
2MASS J17583455--2123215 & $0^{\prime\prime}.1$ & $J = 11.38\pm 0.04$ & $H = 9.53\pm 0.03$  & $K_{s} = 8.44\pm 0.03$\\
DENIS J175834.5--212321  & $0^{\prime\prime}.3$ & $I = 15.26\pm 0.05$ & $J = 11.25\pm 0.06$ & $K_{s} = 8.38\pm 0.05$\\ \hline
\end{tabular}
\end{center}
\end{minipage}
\end{table*}

{\bf IGR J17404--3655:}  A {\em Swift} position was previously obtained for this source 
\citep{landi1539}, leading to a likely identification with a USNO-B1.0 source.  From
follow-up optical spectroscopy that showed a strong H$\alpha$ emission line but no 
other strong emission or absorption line features, \cite{masetti7} identify the 
source as an X-ray binary.  The position of CXOU J174026.8--365537 is consistent with 
the X-ray binary, confirming the identification.  However, we are dubious about the
further classification of the source as an LMXB rather than an HMXB.  The {\em Swift}
spectrum showed a power-law index of $\Gamma\sim 0.24$ \citep{landi1539}, and 
{\em Chandra} shows an even harder spectrum with $\Gamma$ = --0.30$^{+0.30}_{-0.24}$,
which would be very unusual for typical LMXBs and is more commonly seen for HMXBs
that harbor more highly magnetized neutron stars.

{\bf IGR J17507--2647:}  This source was discovered within $\sim$$2.5^{\circ}$ of the 
Galactic center by {\em INTEGRAL} \citep{bird06}, but no further information about this 
source has been reported until now.  The {\em Chandra} energy spectrum for 
CXOU J175039.4--264436 exhibits a high column density with $N_{\rm H} = 
(1.34^{+0.78}_{-0.55})\times 10^{23}$ cm$^{-2}$.  Although this may indicate that the
source is obscured (e.g., an obscured HMXB), Table~\ref{tab:spectra} indicates that
$N_{\rm H_{2}}$ is very high along this line-of-sight, $4.7\times 10^{22}$ cm$^{-2}$, 
and the upper limit on $N_{\rm H,local}$ indicates that the measured absorption may be 
interstellar.  CXOU J175039.4--264436 does not have 2MASS, DENIS, USNO-B1.0, or 
USNO-A2.0 counterparts (also see Figure~\ref{fig:k06}d), providing further evidence 
that the absorption is interstellar.  Although we cannot say definitively what the 
nature of the source is, the result that there is a high level of interstellar 
absorption also indicates that the source is distant, possibly close to the Galactic 
center region at $\sim$8.5~kpc.  This would argue for a luminosity of 
$\sim$$4\times 10^{34}$ ergs~s$^{-1}$, which would suggest that the source is an 
X-ray binary rather than a CV. 

{\bf IGR J17586--2129:}  This source was discovered by {\em INTEGRAL} \citep{bird06}, 
but no further information about this source has been reported until now.  As shown
in Table~\ref{tab:spectra}, CXOU J175834.5--212321 has $N_{\rm H,local}$ in the range
(9--22)$\times 10^{22}$ cm$^{-2}$, providing excellent evidence that this source has 
local absorption.  The position of the {\em Chandra} source is within $0^{\prime\prime}.1$ 
of 2MASS J17583455--2123215, which has IR magnitudes of $J = 11.38\pm 0.04$, 
$H = 9.53\pm 0.03$, and $K_{s} = 8.44\pm 0.03$, and it is within $0^{\prime\prime}.3$ of 
DENIS J175834.5-212321, which has similar IR magnitudes and an $I$-band magnitude of 
$15.26\pm 0.05$ (see Table~\ref{tab:oir}).  The source is not present in the USNO 
catalogs.  The magnitudes clearly indicate a highly reddened source, suggesting a 
large distance (likely at least several kpc).  The 0.3--10 keV flux (see 
Table~\ref{tab:spectra}) indicates a luminosity of $3\times 10^{34}$ ergs~s$^{-1}$ 
for a fiducial distance of 5~kpc, which is too bright for the source to be a CV.  
Also, the hard X-ray spectrum, $\Gamma = 0.23^{+0.59}_{-0.54}$, suggests an HMXB rather 
than an AGN.  Overall, this source is an excellent candidate for being one of the 
obscured IGR HMXBs.

{\bf IGR J18134--1636:}  This source was discovered by {\em INTEGRAL} \citep{bird06}, 
but no further information about this source has been reported until now.  The 
{\em Chandra} spectrum for CXOU J181328.0--163548 shows evidence for local absorption
with $N_{\rm H,local}$ in the range (4--17)$\times 10^{22}$ cm$^{-2}$.  There are no 
convincing optical or IR counterparts to the {\em Chandra} source in the 2MASS, 
DENIS, or USNO catalogs.  The nearest source in these 4 catalogs is a USNO-B1.0 
source that is $1^{\prime\prime}.24$ away from CXOU J181328.0--163548; however, this 
is too far away to claim an association.  The IR source shown in Figure~\ref{fig:k06}e
that appears to be consistent with the {\em Chandra} position is not listed in the
2MASS catalog, and this may be the IR counterpart.  The high column density makes it 
unlikely that this source is a CV.  However, with a power-law index of 
$\Gamma = 1.4^{+0.9}_{-0.8}$, it could either be an X-ray binary or an AGN.

{\bf IGR J18173--2509:}  The position of CXOU J181722.1--250842 is consistent
with the {\em Swift} position for this IGR source reported in \cite{landi1437}.
An optical source in the {\em Swift} error circle was followed-up with optical
spectroscopy, showing that the source is a CV at a distance of $\sim$330 pc 
\citep{masetti7}.  The 0.3--10 keV flux from the {\em Chandra} spectrum of
$8.5\times 10^{-12}$ ergs~cm$^{-2}$~s$^{-1}$ is similar to the 2--10 keV value
quoted for the {\em Swift} spectrum of $1.3\times 10^{-11}$ ergs~cm$^{-2}$~s$^{-1}$.
At a distance of 330 pc, this flux corresponds to a luminosity of $\sim$$10^{32}$ 
ergs~s$^{-1}$, typical of CVs.  However, we find that the {\em Chandra} spectrum 
is well-described by a single power-law in contrast to the two power-law model 
described for the {\em Swift} spectrum by \cite{landi1437}.

{\bf IGR J18308--1232:}  This source was previously identified with the 
{\em XMM-Newton} slew source XMMLS1 J183049.6--123218 \citep{iks08}, and the 
positions of CXOU J183049.9--123219 and the {\em XMM-Newton} source are 
consistent.  Follow-up optical spectroscopy has already been obtained for
the correct optical counterpart, USNO-B1.0 0774-0551687, indicating that
the source is a CV \citep{parisi08,masetti7,butler09}.  The {\em Chandra}
flux level and energy spectrum are significantly different from the 
{\em XMM-Newton} measurements.  The unabsorbed 0.3--10 keV {\em Chandra}
flux level is about 4 times higher than reported in \cite{iks08} for 
{\em XMM-Newton}.  Also, the {\em Chandra} photon index is 
$\Gamma = 0.54^{+0.18}_{-0.31}$, which is significantly harder than the
value of $\Gamma = 1.7$ reported for {\em XMM-Newton}.  At the estimated
distance to the source of 200--300~pc \citep{masetti7,butler09}, the 
{\em Chandra} flux level corresponds to an X-ray luminosity of 
$\sim$$10^{32}$ ergs~s$^{-1}$, which is typical for CVs.

{\bf IGR J18485--0047:}  The position of CXOU J184825.4--004635 is consistent
with the {\em Swift} position for this IGR source reported in \cite{landi1322}.
The {\em Chandra} position does not correspond to any cataloged 2MASS, DENIS,
USNO-B1.0, or USNO-A2.0 sources, but the $K_{s}$-band image shown in 
Figure~\ref{fig:k06}f suggests that a faint IR source may be associated with
CXOU J184825.4--004635.  Also, \cite{landi1322} noted a possible association
between IGR J18485--0047 and the radio source GPSR5 31.897+0.317 \citep{becker94}.
The {\em Chandra} source is within $1^{\prime\prime}.1$ of the radio source, and
the 90\% confidence error radius on the radio source is $2^{\prime\prime}.2$, so
the association is likely.  The flux of the radio source is 8.9 mJy at 5 GHz
and 21 mJy at 1.4 GHz, and it is consistent with being a point source.  As for
the X-ray properties, both the {\em Swift} and {\em Chandra} spectra show 
evidence for absorption well in excess of the Galactic value.  The association
with a radio source may indicate an AGN nature for the source; however, CVs
and X-ray binaries can also be radio emitters, so the classification is not
certain.  If the source is an AGN, then the high levels of absorption would 
indicate a Seyfert 2 classification.

{\bf IGR J19267+1325:} The {\em Chandra} information for CXOU J192626.9+132205
was reported previously in \cite{tomsick1649}, and the {\em Chandra} and {\em Swift} 
results for this source are discussed in detail in \cite{rtc09}.  After
the identification of the optical counterpart \citep{tomsick1649}, follow-up
optical spectroscopy showed that the source is a CV 
\citep{steeghs1653,masetti7,butler09}.  Furthermore, an X-ray period of 938.6~s
has been measured and is identified as the white dwarf spin period, indicating
that the source is a CV of Intermediate Polar type \citep{evans1669}.  A second 
period may also be present at 4.58 hours, and this may be the binary orbital
period \citep{evans1669}.  Somewhat discrepant distances of 580~pc 
\citep{masetti7} and 250~pc \citep{butler09} have been derived.  For this range
of distances, the X-ray flux that we measure indicates a range of 0.3--10 keV
luminosities of $7\times 10^{31}$ to $4\times 10^{32}$ ergs~s$^{-1}$.

{\bf IGR J19443+2117:}  Based on a {\em Swift} detection of this source, 
\cite{landi09} suggest associations with the radio source NVSS J194356+211826
and the IR source 2MASS J19435624+2118233.  We confirm these associations as
CXOU J194356.2+211823 is within $0^{\prime\prime}.3$ of the 2MASS source.  
\cite{landi09} discuss the nature of the source, which is a relatively 
bright radio source (103 mJy at 1.4 GHz) and also has a $K_{s}$-magnitude
of $13.98\pm 0.07$.  The {\em Chandra} spectrum and flux are similar to 
those reported by \cite{landi09}, but we find even stronger evidence that
the column density is in excess of the Galactic value.  Overall, we agree
with the conclusion of \cite{landi09} that IGR J19443+2117 is most likely an 
AGN, but an optical or IR spectral confirmation would strengthen this 
conclusion.

\section{Summary and Conclusions}

The main goal of this {\em Chandra} project has been to localize and measure
the soft X-ray spectra of IGR sources in the Galactic plane in order to 
determine their nature or to guide follow-up observations to determine their
nature.  In this paper, we have reported results from 22 {\em Chandra} 
observations, and we have identified likely soft X-ray counterparts in 18
cases.  Here, we summarize the different types of sources that we have found.

{\bf Two X-ray binaries and five X-ray binary candidates:}  We confirm that
IGR~J14331--6112 is an HMXB based on its positional coincidence with the 
previously suggested counterpart \citep{masetti6}, and we show that this 
source has a very strong iron K$\alpha$ emission line, which is similar to
what has been seen for some of the other IGR HMXBs.  We also confirm the
\cite{masetti7} optical identification of IGR~J17404--3655.  While we
agree that the source is an X-ray binary, we argue that the hard X-ray 
spectrum may indicate that the source is an HMXB rather than an LMXB.
Hard X-ray spectra ($\Gamma < 1$) combined with large optical/IR extinctions
(and thus, large distances) argue for an HMXB nature, and we suggest that 
IGR~J16287--5021, IGR~J17354--3255, IGR~J17507--2647, and IGR J17586--2129 
may be HMXBs based on this evidence.  Finally, comparing {\em Swift} and
{\em Chandra} spectra shows that the column density for IGR~J13186--6257
is high and variable, which suggests the possibility that the source is 
and HMXB.  

{\bf Three CVs and one CV candidate:}  For IGR~J19267+1325, the {\em Chandra}
position led to the identification of the optical counterpart that was
shown to be a CV.  In addition, we confirm the previously suggested
optical identifications for IGR~J18173--2509 and IGR~J18308--1232, 
and these are also CVs.  We also mention the new possibility that
IGR~J15529--5029 may be a CV based on a hard X-ray spectrum 
($\Gamma < 1$) combined with relatively low optical/IR extinctions.

{\bf One AGN and two AGN candidates:}  We confirm that IGR~J14471--6414 is 
an AGN based on its positional coincidence with the previously suggested 
counterpart, which is a Seyfert 1.2 at a redshift of $z = 0.053$
\citep{masetti6}.  We confirm the association of IGR~19443+2117 with
2MASS and radio counterparts \citep{landi09}, but spectral confirmation
is still required to confirm that this source is an AGN.  Finally, 
we identify IGR~J18485--0047 with a strong radio source.  Based on 
the radio emission and the strong X-ray absorption, we suggest that 
it may be a Seyfert 2 AGN.

{\bf Two SNRs:}  IGR~J14003--6326 and IGR~J17448--3232 are both circular
extended sources, indicative of being SNRs.  In the former case, the 
putative pulsar appears to be embedded in the SNR, and the hard X-ray
emission may be dominated by the PWN.  In the latter case, a hard
point-like source at the edge of the SNR may indicate that the pulsar
was kicked when the explosion occurred.  Prior to this work, only 
1 IGR source was identified as an SNR, IGR~J18135-1751, which has 
also been detected at TeV energies \citep{ubertini05}.  The {\em INTEGRAL} 
discovery of 2 more SNRs may effect estimates of the hard X-ray emission 
levels from SNRs in general, and may have implications for the numbers of 
SNRs that future hard X-ray satellites, such as {\em NuSTAR} 
\citep{harrison09} will detect.

{\bf Two unclassified sources:}  We have identified IGR J11098--6457 
with CXOU J110926.4--650224 and IGR J18134--1636 with CXOU J181328.0--163548, 
but the nature of these sources is unclear.

{\bf Sources lacking clear counterparts:}  We did not find clear 
{\em Chandra} counterparts for IGR J07295--1329, IGR J09485--4726, 
IGR J17461--2204, or IGR J17487--3124.  This may indicate that these 
are transient or variable X-ray sources.

\acknowledgments

JAT acknowledges partial support from {\em Chandra} award number
GO8-9055X issued by the {\em Chandra X-Ray Observatory Center}, 
which is operated by the Smithsonian Astrophysical Observatory for 
and on behalf of the National Aeronautics and Space Administration 
(NASA), under contract NAS8-03060.  JAT would like to thank 
Michael McCollough for useful conversations about {\em Chandra}
source detection.  This publication makes use of data products from 
the Two Micron All Sky Survey, which is a joint project of the 
University of Massachusetts and the Infrared Processing and Analysis 
Center/California Institute of Technology, funded by NASA and the 
National Science Foundation.  This research makes use of the USNOFS 
Image and Catalogue Archive operated by the United States Naval 
Observatory, Flagstaff Station, the SIMBAD database, operated at 
CDS, Strasbourg, France, and the Deep Near Infrared Survey of the 
Southern Sky (DENIS).

\appendix

The {\em Chandra} source lists for all 22 observations are available
on-line.  The lists include sources from the ACIS-I detector only, and 
the source detection methodology is describe in \S$3.1$.  The tables 
are in the same format as Table~\ref{tab:counterparts} with the CXOU 
{\em Chandra} name for each source, the angular distance from the 
center of the {\em INTEGRAL} error circle, the {\em Chandra} position, 
the number of ACIS counts collected in the 0.3--10 keV energy band, and 
the hardness ratio as defined in \S$3.1$.  In a small number of cases
near the edges of the detector, background subtraction leads to sources
that were detected by {\ttfamily wavdetect} having negative numbers of
counts.  As we know that some of the {\ttfamily wavdetect} sources are
spurious, we removed the sources that have fewer than 1 count.  In 
addition, the tables do not include hardness ratios for sources with
fewer than 5 ACIS counts as the errors are too large in these cases.
The tables for IGR~J07295--1329, IGR~J09485--4726, IGR~J17461--2204, 
and IGR~J17487--3124 may be especially useful as these were the four
sources for which we did not find a clear {\em Chandra} counterpart.
Thus, results from future observations could be compared to the
these tables to look for X-ray variability that could help to identify 
the counterparts of the IGR sources.  We note that the 90\% confidence
{\em INTEGRAL} error circles for the 4 sources listed above have radii
of $5^{\prime}.4$, $4^{\prime}.9$, $3^{\prime}.7$, and $3^{\prime}.0$,
respectively.


\begin{thebibliography}{}

\bibitem[\protect\astroncite{{Abdo}}{2009}]{abdo09}
{Abdo}, A.~A.,  2009, arXiv:0902.1340 [astro-ph]

\bibitem[\protect\astroncite{{Anders} \& {Grevesse}}{1989}]{ag89}
{Anders}, E., \& {Grevesse}, N.,  1989, Geochimica et Cosmochimica Acta, 53,
  197

\bibitem[\protect\astroncite{{Balucinska-Church} \& {McCammon}}{1992}]{bm92}
{Balucinska-Church}, M., \& {McCammon}, D.,  1992, ApJ, 400, 699

\bibitem[\protect\astroncite{{Becker} et~al.}{1994}]{becker94}
{Becker}, R.~H., {White}, R.~L., {Helfand}, D.~J., \& {Zoonematkermani}, S.,
  1994, ApJS, 91, 347

\bibitem[\protect\astroncite{{Bird} et~al.}{2006}]{bird06}
{Bird}, A.~J., et~al., 2006, ApJ, 636, 765

\bibitem[\protect\astroncite{{Bodaghee} et~al.}{2007}]{bodaghee07}
{Bodaghee}, A., et~al., 2007, A\&A, 467, 585

\bibitem[\protect\astroncite{{Bozzo}, {Falanga} \& {Stella}}{2008}]{bfs08}
{Bozzo}, E., {Falanga}, M., \& {Stella}, L.,  2008, ApJ, 683, 1031

\bibitem[\protect\astroncite{{Bulgarelli} et~al.}{2009}]{bulgarelli2017}
{Bulgarelli}, A., et~al., 2009, The Astronomer's Telegram, 2017

\bibitem[\protect\astroncite{{Butler} et~al.}{2009}]{butler09}
{Butler}, S.~C., et~al., 2009, ApJ, 698, 502

\bibitem[\protect\astroncite{{Cash}}{1979}]{cash79}
{Cash}, W.,  1979, ApJ, 228, 939

\bibitem[\protect\astroncite{{Chaty} et~al.}{2008}]{chaty08}
{Chaty}, S., {Rahoui}, F., {Foellmi}, C., {Tomsick}, J.~A., {Rodriguez}, J., \&
  {Walter}, R.,  2008, A\&A, 484, 783

\bibitem[\protect\astroncite{{Dame}, {Hartmann} \& {Thaddeus}}{2001}]{dht01}
{Dame}, T.~M., {Hartmann}, D., \& {Thaddeus}, P.,  2001, ApJ, 547, 792

\bibitem[\protect\astroncite{{Davis}}{2001}]{davis01}
{Davis}, J.~E.,  2001, ApJ, 562, 575

\bibitem[\protect\astroncite{{Ebisawa} et~al.}{2005}]{ebisawa05}
{Ebisawa}, K., et~al., 2005, ApJ, 635, 214

\bibitem[\protect\astroncite{{Evans}, {Beardmore} \&
  {Osborne}}{2008}]{evans1669}
{Evans}, P.~A., {Beardmore}, A.~P., \& {Osborne}, J.~P.,  2008, The
  Astronomer's Telegram, 1669

\bibitem[\protect\astroncite{{Filliatre} \& {Chaty}}{2004}]{fc04}
{Filliatre}, P., \& {Chaty}, S.,  2004, ApJ, 616, 469

\bibitem[\protect\astroncite{{Funk} et~al.}{2007}]{funk07}
{Funk}, S., et~al., 2007, A\&A, 470, 249

\bibitem[\protect\astroncite{{Garmire} et~al.}{2003}]{garmire03}
{Garmire}, G.~P., {Bautz}, M.~W., {Ford}, P.~G., {Nousek}, J.~A., \& {Ricker},
  G.~R.,  2003,
\newblock in X-Ray and Gamma-Ray Telescopes and Instruments for Astronomy.
  Edited by Joachim E. Truemper, Harvey D. Tananbaum. Proceedings of the SPIE,
  4851, 28

\bibitem[\protect\astroncite{{Gehrels}}{1986}]{gehrels86}
{Gehrels}, N.,  1986, ApJ, 303, 336

\bibitem[\protect\astroncite{{Hands} et~al.}{2004}]{hands04}
{Hands}, A.~D.~P., {Warwick}, R.~S., {Watson}, M.~G., \& {Helfand}, D.~J.,
  2004, MNRAS, 351, 31

\bibitem[\protect\astroncite{{Harrison} et~al.}{2009}]{harrison09}
{Harrison}, F., et~al., 2009,
\newblock in American Astronomical Society Meeting Abstracts, Vol. 213,
  \#452.02

\bibitem[\protect\astroncite{{Helfand} et~al.}{2007}]{helfand07}
{Helfand}, D.~J., {Gotthelf}, E.~V., {Halpern}, J.~P., {Camilo}, F., {Semler},
  D.~R., {Becker}, R.~H., \& {White}, R.~L.,  2007, ApJ, 665, 1297

\bibitem[\protect\astroncite{{Ibarra}, {Kuulkers} \& {Saxton}}{2008}]{iks08}
{Ibarra}, A., {Kuulkers}, E., \& {Saxton}, R.,  2008, The Astronomer's
  Telegram, 1397

\bibitem[\protect\astroncite{{in't Zand}}{2005}]{intzand05}
{in't Zand}, J.~J.~M.,  2005, A\&A, 441, L1

\bibitem[\protect\astroncite{{Kalberla} et~al.}{2005}]{kalberla05}
{Kalberla}, P.~M.~W., {Burton}, W.~B., {Hartmann}, D., {Arnal}, E.~M.,
  {Bajaja}, E., {Morras}, R., \& {P{\"o}ppel}, W.~G.~L.,  2005, A\&A, 440, 775

\bibitem[\protect\astroncite{{Keek}, {Kuiper} \& {Hermsen}}{2006}]{keek06}
{Keek}, S., {Kuiper}, L., \& {Hermsen}, W.,  2006, The Astronomer's Telegram,
  810

\bibitem[\protect\astroncite{{Kuulkers} et~al.}{2006}]{kuulkers874}
{Kuulkers}, E., et~al., 2006, The Astronomer's Telegram, 874, 1

\bibitem[\protect\astroncite{{Landi} et~al.}{2007a}]{landi1273}
{Landi}, R., et~al., 2007a, The Astronomer's Telegram, 1273

\bibitem[\protect\astroncite{{Landi} et~al.}{2008a}]{landi1437}
{Landi}, R., {Masetti}, N., {Malizia}, A., {Bazzano}, A., {Fiocchi}, M.,
  {Bird}, A.~J., \& {Dean}, A.~J.,  2008a, The Astronomer's Telegram, 1437

\bibitem[\protect\astroncite{{Landi} et~al.}{2008b}]{landi1538}
{Landi}, R., et~al., 2008b, The Astronomer's Telegram, 1538

\bibitem[\protect\astroncite{{Landi} et~al.}{2008c}]{landi1539}
{Landi}, R., et~al., 2008c, The Astronomer's Telegram, 1539

\bibitem[\protect\astroncite{{Landi} et~al.}{2007b}]{landi1323}
{Landi}, R., et~al., 2007b, The Astronomer's Telegram, 1323

\bibitem[\protect\astroncite{{Landi} et~al.}{2007c}]{landi1322}
{Landi}, R., et~al., 2007c, The Astronomer's Telegram, 1322

\bibitem[\protect\astroncite{{Landi} et~al.}{2009}]{landi09}
{Landi}, R., et~al., 2009, A\&A, 493, 893

\bibitem[\protect\astroncite{{Lebrun} et~al.}{2003}]{lebrun03}
{Lebrun}, F., et~al., 2003, A\&A, 411, L141

\bibitem[\protect\astroncite{{Lommen} et~al.}{2005}]{lommen05}
{Lommen}, D., {Yungelson}, L., {van den Heuvel}, E., {Nelemans}, G., \&
  {Portegies Zwart}, S.,  2005, A\&A, 443, 231

\bibitem[\protect\astroncite{{Malizia} et~al.}{2007}]{malizia07}
{Malizia}, A., et~al., 2007, ApJ, 668, 81

\bibitem[\protect\astroncite{{Masetti} et~al.}{2008}]{masetti6}
{Masetti}, N., et~al., 2008, A\&A, 482, 113

\bibitem[\protect\astroncite{{Masetti} et~al.}{2009}]{masetti7}
{Masetti}, N., et~al., 2009, A\&A, 495, 121

\bibitem[\protect\astroncite{{Matt} \& {Guainazzi}}{2003}]{mg03}
{Matt}, G., \& {Guainazzi}, M.,  2003, MNRAS, 341, L13

\bibitem[\protect\astroncite{{Negueruela} et~al.}{2006}]{negueruela06}
{Negueruela}, I., {Smith}, D.~M., {Reig}, P., {Chaty}, S., \& {Torrej{\'o}n},
  J.~M.,  2006,
\newblock in ESA SP-604: The X-ray Universe 2005, ed. A. {Wilson}, 165

\bibitem[\protect\astroncite{{Parisi} et~al.}{2008}]{parisi08}
{Parisi}, P., {Masetti}, N., {Jimenez}, E., {Chavushyan}, V., {Bassani}, L.,
  {Bazzano}, A., \& {Bird}, A.~J.,  2008, The Astronomer's Telegram, 1710

\bibitem[\protect\astroncite{{Rahoui} et~al.}{2008}]{rahoui08}
{Rahoui}, F., {Chaty}, S., {Lagage}, P.-O., \& {Pantin}, E.,  2008, A\&A, 484,
  801

\bibitem[\protect\astroncite{{Rodriguez}, {Tomsick} \& {Chaty}}{2009}]{rtc09}
{Rodriguez}, J., {Tomsick}, J.~A., \& {Chaty}, S.,  2009, A\&A, 494, 417

\bibitem[\protect\astroncite{{Sguera} et~al.}{2006}]{sguera06}
{Sguera}, V., et~al., 2006, ApJ, 646, 452

\bibitem[\protect\astroncite{{Steeghs} et~al.}{2008}]{steeghs1653}
{Steeghs}, D., {Knigge}, C., {Drew}, J., {Unruh}, Y., \& {Greimel}, R.,  2008,
  The Astronomer's Telegram, 1653

\bibitem[\protect\astroncite{{Sugizaki} et~al.}{2001}]{sugizaki01}
{Sugizaki}, M., {Mitsuda}, K., {Kaneda}, H., {Matsuzaki}, K., {Yamauchi}, S.,
  \& {Koyama}, K.,  2001, ApJS, 134, 77

\bibitem[\protect\astroncite{{Tomsick}}{2009}]{tomsick2022}
{Tomsick}, J.~A.,  2009, The Astronomer's Telegram, 2022

\bibitem[\protect\astroncite{{Tomsick} et~al.}{2006}]{tomsick06}
{Tomsick}, J.~A., {Chaty}, S., {Rodriguez}, J., {Foschini}, L., {Walter}, R.,
  \& {Kaaret}, P.,  2006, ApJ, 647, 1309

\bibitem[\protect\astroncite{{Tomsick} et~al.}{2008a}]{tomsick08a}
{Tomsick}, J.~A., {Chaty}, S., {Rodriguez}, J., {Walter}, R., \& {Kaaret}, P.,
  2008a, \apj, 685, 1143

\bibitem[\protect\astroncite{{Tomsick} et~al.}{2009}]{tomsick09}
{Tomsick}, J.~A., {Chaty}, S., {Rodriguez}, J., {Walter}, R., {Kaaret}, P., \&
  {Tovmassian}, G.,  2009, ApJ, 694, 344

\bibitem[\protect\astroncite{{Tomsick} et~al.}{2008b}]{tomsick08}
{Tomsick}, J.~A., et~al., 2008b, ApJ, 680, 593

\bibitem[\protect\astroncite{{Tomsick} et~al.}{2008c}]{tomsick1649}
{Tomsick}, J.~A., {Rodriguez}, J., {Chaty}, S., {Walter}, R., \& {Kaaret}, P.,
  2008c, The Astronomer's Telegram, 1649

\bibitem[\protect\astroncite{{Ubertini} et~al.}{2005}]{ubertini05}
{Ubertini}, P., et~al., 2005, ApJ, 629, L109

\bibitem[\protect\astroncite{{Ubertini} et~al.}{2003}]{ubertini03}
{Ubertini}, P., et~al., 2003, A\&A, 411, L131

\bibitem[\protect\astroncite{{Vercellone} et~al.}{2009}]{vercellone2019}
{Vercellone}, S., et~al., 2009, The Astronomer's Telegram, 2019

\bibitem[\protect\astroncite{{Walter} et~al.}{2003}]{walter03}
{Walter}, R., et~al., 2003, A\&A, 411, L427

\bibitem[\protect\astroncite{{Walter} \& {Zurita Heras}}{2007}]{wz07}
{Walter}, R., \& {Zurita Heras}, J.,  2007, A\&A, 476, 335

\bibitem[\protect\astroncite{{Walter} et~al.}{2006}]{walter06}
{Walter}, R., et~al., 2006, A\&A, 453, 133

\bibitem[\protect\astroncite{{Weisskopf}}{2005}]{weisskopf05}
{Weisskopf}, M.~C.,  2005, arXiv:astro-ph/0503091

\bibitem[\protect\astroncite{{Wilms}, {Allen} \& {McCray}}{2000}]{wam00}
{Wilms}, J., {Allen}, A., \& {McCray}, R.,  2000, ApJ, 542, 914

\bibitem[\protect\astroncite{{Winkler} et~al.}{2003}]{winkler03}
{Winkler}, C., et~al., 2003, A\&A, 411, L1

\end{thebibliography}

\end{document}